\documentclass[preprint,11pt]{elsarticle}

\usepackage{graphicx}
\usepackage{multirow}
\usepackage{amssymb}
\usepackage{float}
\usepackage{amsmath}
\usepackage{xcolor}
\usepackage{natbib}
\usepackage{setspace}
\newcommand{\Rho}{\mathrm{P}}
\usepackage{geometry}
\usepackage{mathrsfs}
\usepackage[draft]{changes}
\journal{}
\begin{document}

\begin{frontmatter}



\title{A tensor-based unified approach for clustering coefficients in financial multiplex networks}

	\author[Bic1]{Paolo Bartesaghi}
		\author[Catt]{Gian Paolo Clemente}
		\author[Bic1]{Rosanna Grassi\corref{cor1}}
		
		\cortext[cor1]{\emph{Corresponding author. email: rosanna.grassi@unimib.it}} 
		\address[Bic1]{University of Milano - Bicocca, Via Bicocca degli Arcimboldi 8, 20126 Milano, Italy, email: paolo.bartesaghi@unimib.it; rosanna.grassi@unimib.it}
		\address[Catt]{Universit\`{a} Cattolica del Sacro Cuore di Milano, Largo Gemelli 1, 20123 Milano, Italy email: gianpaolo.clemente@unicatt.it}


\begin{abstract}
Big data and the use of advanced technologies are relevant topics in the financial market. In this context, complex networks became extremely useful in describing the structure of complex financial systems. In particular, the time evolution property of the stock markets have been described by temporal networks. However, these approaches fail to consider the interactions over time between assets. To overcome this drawback, financial markets can be described by multiplex networks where the different relations between nodes can be conveniently expressed structuring the network through different layers. To catch this kind of interconnections we provide new local clustering coefficients for multiplex networks, looking at the network from different perspectives depending on the node position, as well as a global clustering coefficient for the whole network. We also prove that all the well-known expressions for clustering coefficients existing in the literature, suitably extended to the multiplex framework, may be unified into our proposal. By means of an application to the multiplex temporal financial network, based on the returns of the S\&P100 assets, we show that the proposed measures prove to be effective in describing dependencies between assets over time. 

\end{abstract}


\end{frontmatter}


\section{Introduction}
\label{Introduction}

Nowadays, a huge amount of multidimensional heterogeneous data are collected from economic, social, financial and technological fields. In particular, big data and the use of advanced technologies are relevant topics in the financial market. The digital transformation is pushing organizations to move towards more data-driven business and innovation models. 
On the one hand, financial analysts use external and alternative data to make investment decisions.
On the other hand, financial industries use big data through different predictive methods and monitor various spending patterns to develop large decision-making models. Hence, big data is receiving more attention in the financial framework, where information affects success and
production factors, enhancing and consolidating our
understanding of financial markets. 
In this context, a useful tool to manage and analyse such large amount of data is offered in the literature by complex networks (see, e.g., \cite{Boccaletti2014,Hyland2021}). Indeed, the description and integration of the structure of financial complex systems can be efficiently encoded in a suitable network mathematical model, providing  a better comprehension of topological features of these systems and the related dynamical processes (see, e.g., \cite{Dehmer, WangSpill}).\\
Although the research related to big data and financial issues is extremely new (\cite{Hasan}), complex networks have been usefully applied in finance for several purposes (see, e.g., \cite{Chinazzi} and \cite{Nagurney}). Prominent examples are, for instance, the analysis of the interconnected nature of financial systems (\cite{Battiston2018}, \cite{Barja}), the modelling of the dynamics of financial contagion and systemic vulnerabilities within a system (\cite{Caccioli2018}), the analysis of stock markets via temporal networks (see, e.g., \cite{ZHAO}). 
In particular, the correlation-based network has become an effective tool to investigate the correlation between complex financial systems (\cite{Mantegna1999}) and to catch the dependence between underlying random variables \citep{Pena2013}.
The typical approach is to study these data by means of a sequence of correlation-based monoplex networks, one for each time period (see, e.g., \cite{Onnela2003b} and \cite{pozzi13}). However, these approaches do not consider the presence of autocorrelation in financial markets. The  serial  correlation  in  stock  returns  is indeed a central issue in different financial areas (see, e.g., \cite{English}, \cite{Cont2007}). Although contrasting results have been obtained in this area showing different behaviours for short-term and long-term analyses (see \cite{Kadlec}, \cite{Sias}, \cite{Lo}, \cite{Fama}), it is interesting to introduce the effect of correlation over time in a network context. \\
To this reason and to catch the coexistence of multiple types of interactions within interconnected systems, we focus on the study of the multilayer nature of real-world networks. This kind of networks are currently classified in different ways, including heterogeneous information networks (\cite{Chen2021}), multilayer networks (\cite{DeDomenico2013,Kivela2014,Zhang2017,Scabini2019}) and multidimensional networks (\cite{Berlingerio2013}). In this work we will refer specifically to multiplex networks, that is networks in which the nodes in every layer are exactly the same.

Beyond this classification, all these networks encode in the same complex structure relations of different nature and with different meaning. Indeed, it is worth pointing out that networks usually have various features in different layers that cannot be properly highlighted aggregating the layers by the overlay or the projection operation (see, e.g., \cite{Battiston2014}).  Additionally, the meaning of interlinks connecting nodes in different layers goes over the simple formal representation.
These aspects motivate a study of the complex object represented by multiplex networks preserving as much as possible the original structure, even using a more sophisticated mathematical tool (\cite{Lv2021,Park2016}).

In this framework, special attention is paid to the  clustering coefficient, an important measure of network topology. This indicator measures the degree to which nodes in a graph tend to cluster together and it is widely used in the financial literature (see, e.g., \cite{Minoiu} and \cite{Tabak2014}) \\
Alternative definitions of clustering coefficient in multidimensional networks have been proposed (see, in particular, \cite{DeDomenico2013,Cozzo2015}) and all of them revealed that a univocal extension is not possible in a multiplex context.
In order to gain a deeper insight into specific multiplex financial networks, we propose new local clustering coefficients for weighted undirected multiplex networks with non-diagonal couplings. In doing so, we look at the network from different perspectives, depending on the node position. Indeed, a node can be well-clustered into the level to which it belongs, or transversely across the levels, or even globally in the whole system. Therefore, we provide alternative coefficients that catch the various situations, capturing all these different features intrinsically related to the complex topology of the multiplex network.\\  
We formalize these definitions by using tensors. Indeed, from a mathematical point of view, tensors offer an effective and elegant representation for generalizing relations between nodes in the multiplex network context (see \cite{DeDomenico2013,wang2017,wang2020}). We prove that the well-known formulas for the clustering coefficients already existing in the literature for weighted monoplex networks (see \cite{Barrat2004} and \cite{Onnela2005}), suitably extended to the multiplex case, may be unified into our proposal, both in terms of tensors and supradjacency matrices.

The proposed measures prove to be effective in catching the multiplex nature of temporal financial networks. In particular, we focus on the returns of the $102$ leading U.S. stocks constituents of the S\&P 100 index for the time-period ranging from January 2001 to June 2017. To this end, we build a multiplex network, where each layer considers the correlations between assets at a specific time period, while inter-layer links consider the correlation over time. This specific structure allows to include in the analysis the possible presence of autocorrelation between stock returns over time. Results show how the pattern of clustering coefficients is consistent with main financial events that characterized the analysed period. Additionally, we show how local coefficients allow to emphasize peculiar behaviours at both sector and asset level.

The paper is organized as follows. In section \ref{Mathematics of multiplex networks} we introduce the mathematical formalism needed for dealing with multiplex networks. In section \ref{Triangles in multiplex networks} we define the definition of triangles in a multidimensional framework. In section \ref{Clustering for multiplex networks} we provide the general formulation of the local clustering coefficients. Formulas are defined by using both tensor and supradjacency matrices. Also, we show how the proposed coefficients are generalisations of the classical definitions provided in the literature for monoplex networks. In section \ref{toyex} the different meanings of the proposed coefficients are emphasized by means of a simple example. In section \ref{empar} we apply our proposal to a real multiplex network based on S\&P 100 assets. Conclusions follow.

\subsection{Related Literature on clustering coefficients}

There is a wide literature about clustering coefficients in complex networks. A first distinction regards global and local coefficients. In addition to this, a large number of coefficients has been proposed for undirected and directed, unweighted and weighted, monoplex and multiplex networks. For this reason, we shortly remind here the main contributions in this field focusing on weighted, undirected monoplex, multilayer and multiplex networks. \\ 
Different definitions of clustering coefficient have been proposed for monoplex networks, at a global (see \cite{Newman} and \cite{WasFaust}) and at a local level (\cite{Watts_1998}). Since then, alternative attempts of generalisation have been considered by several authors in different directions. 
Two well-known local coefficients for weighted networks have been introduced in \cite{Barrat2004, Onnela2005}, which differ in the weight assigned to the triangles involved. Other novel coefficients can be found in \cite{Zhang2005,Opsahl}. The authors in \cite{Saram} present a comparative study of the various coefficients introduced in the literature, stressing advantages and limitations.\\
Very few attempts refer to multilayer networks. A complete formal and systematic approach can be found in \cite{DeDomenico2013} where the authors propose a tensorial representation of multilayer networks. Many topological indicators are extended to this context and, among them, a global clustering coefficient is defined.\\
Several researchers focused on the extension of these coefficients to multiplex networks (\cite{Cozzo2015,Battiston2014,criado2012}). In a general context of structural measures, two clustering coefficients are proposed in \cite{Battiston2014} for node-aligned multiplex networks. These coefficients depend on a generalised definition of triads and triangles (\cite{Baxter2016}). In the same vein, in \cite{Cozzo2015} the transitivity is explored in multiplex networks, generalising both triad relations and clustering coefficients.  The two local clustering coefficients proposed by \cite{criado2012} are based on two alternative definitions of the neighbourhood of a node.  \\
However, to the best of our knowledge, definitions of local clustering coefficients for multiplex networks with non-diagonal couplings are not available in the literature.

In the present paper, we propose a novel mathematical tool which is aimed at describing a node-aligned multiplex network where a node can be connected not only to its counterpart but also to different nodes in other layers. Unlike the classical multiplex networks in the literature (as for instance in \cite{Nicosia2015}), there is a wide class of node-aligned networks where an \lq\lq explicit cost\rq\rq can be associated  with interlayer connections among replicas of the same node. Hence, we refer to a definition of triangles in which the three sides have the same nature, regardless of whether they connect counterparts of the same node on different levels or different nodes.

\section{Mathematics of multiplex networks}
\label{Mathematics of multiplex networks}

Formally, a network is represented by a graph $G=(V,E)$, where $V$ is the set of $N$ nodes and $E\subseteq V\times V$ the set of edges (or links).
Two nodes are adjacent if there is an edge $(i,j)$ connecting them. 
We consider simple graphs, i.e without loops and multiple edges. An undirected graph is a graph in which if $(i,j)\in E,$ then $(j,i)\in E$.
The adjacency relations between pairs of nodes can be conveniently represented by a $N$-square symmetric matrix $\textbf{A}$ (the adjacency matrix) with entries $a_{ij}=1$ if $(i,j) \in E$, $0$ otherwise. \\ 
If any edge $(i,j)\in E$ is associated with a positive real number $w_{ij}$, then both the edges and the graph are weighted. We set that $w_{ij}=0$ if and only if $(i,j)\notin E$, then 
the real $N$-square matrix $\textbf{W}$ (the weighted adjacency matrix)  with entries $w_{ij}$ completely describes $G$. In particular, if $w_{ij}=1$ for all edges $(i,j)\in E$, then $\textbf{W}=\textbf{A}$ and $G$ is called unweighted. Introducing weights, $G$ is undirected if, when $(i,j)\in E$, then $w_{ij}=w_{ji}$. This implies that the matrix $\textbf{W}$ is symmetric.\\
In this paper we refer to multiplex networks that are node-aligned, with non-diagonal couplings (see \cite{Kivela2014}). Specifically, a multiplex network consists of a family of networks $G_a=(V_a,E_a),a=1,...,L$ where each network $G_a=(V_a,E_a)$ is located in a layer $a$ and a node $i \in V_a$ is adjacent to $j \in V_b$, $\forall a,b=1,...,L$ if there is an edge connecting them. Note that, in node-aligned networks, all nodes are shared between all layers, namely, $V_a=V_b=V,\ \forall a,b=1,...,L$, and
that $E_a$ collects all the edges connecting nodes on layer $a$ to nodes within the same layer (intra-layer connections) and to nodes on different layers (inter-layer connections).
We assume that the network is non-diagonal coupled, that is inter-layer edges may exist not only between nodes and their counterparts, but links between a node $i$ in a given layer and a node $j \neq i$ in a different layer are allowed. From now on, we will refer to this kind of networks briefly as multiplex network.

We adopt the tensor formalism for general multilayer networks, described in \cite{DeDomenico2013}. In order to facilitate the reader in associating indices to the corresponding objects, we introduce the following notations. We use Latin letters $i,j,k,..$ and $a,b,c,...$ for objects, namely nodes and levels respectively; and Greek letters $\mu, \nu, \rho, \sigma, \dots$ and $\alpha, \beta, \gamma, \delta, \dots$ for components of vectors or tensors, again, for nodes and levels, respectively. In particular, $v_{\nu}(i)$ represents the $\nu$ component of a general covariant vector on node $i$ and $v^{\mu}(i)$ the $\mu$ component of a general contravariant vector on the same node. In particular, $e^{\mu}(i)$ are the components of the standard basis in ${\mathbb R}^N$, equal to $1$ if $\mu =i$ and $0$ otherwise. We denote by:
\begin{itemize}
	\item $E^{\mu}_{\nu}(i,j)=e^{\mu}(i)e_{\nu}(j)$ the second order tensor canonical basis in ${\mathbb R}^{N\times N}$. This tensor is represented by a $N$- square matrix where the $(i,j)$-entry is 1, 0 otherwise.
	\item $E^{\alpha}_{\beta}(a,b)=e^{\alpha}(a)e_{\beta}(b)$ the second order tensor canonical basis in ${\mathbb R}^{L\times L}$. This tensor is represented by a $L$- square matrix where the $(a,b)$-entry is 1, 0 otherwise.
	\item $E^{\mu \alpha}(i,a)=e^{\mu}(i)e^{\alpha}(a)$ the second-order tensor canonical basis in ${\mathbb R}^{N\times L }$. This tensor is represented by a $N\times L$ matrix where the $(i,a)$-entry is 1, 0 otherwise.
	\item $E^{\mu \alpha}_{\nu \beta}(i,j;a,b)=e^{\mu}(i)e_{\nu}(j)e^{\alpha}(a)e_{\beta}(b)$ the fourth-order tensor canonical basis in ${\mathbb R}^{N\times N\times L \times L}$, where the $(i,j;a,b)$-entry is 1, 0 otherwise.
\end{itemize}

We denote by $W^\mu_\nu(a,b)$ the second order inter-layer adjacency tensor for nodes on layers $a$ and $b$. It can be expressed as $W^\mu_\nu(a,b)=\sum_{i,j=1}^{N}w_{ij}(a,b)E^{\mu}_{\nu}(i,j)$, where $w_{ij}(a,b)$ represents the weight of the link between node $i$ on level $a$ and node $j$ on level $b$. If we focus on the intra-layer connections, i.e. if $a=b$, we set $W^\mu_\nu(a,a)=W^\mu_\nu(a)$, which corresponds to the weighted adjacency matrix of order $N$ of a monoplex network. The multiplex adjacency tensor $M$, that belongs to ${\mathbb R}^{N\times N\times L \times L}$, is therefore expressed as $M^{\mu \alpha}_{\nu \beta}=\sum_{a,b=1}^{L}W^\mu_\nu(a,b)E^{\alpha}_{\beta}(a,b)$.

Notice that $M^{\mu \alpha}_{\nu \beta}$ is a fourth order tensor, encoding all the existing relations between nodes across all layers. 
Let $A^{\mu \alpha}_{\nu \beta}$ be the binary adjacency tensor, obtained by setting all non-zero weights in $M^{\mu \alpha}_{\nu \beta}$ equal to $1$. In view of the subsequent computation of the number and weight of potential triangles, we need a proper definition of the complete multiplex network. A complete multiplex network is meant to be described by the adjacency tensor $F^{\mu \alpha}_{\nu \beta}=U^{\mu \alpha}_{\nu \beta}-I^{\mu \alpha}_{\nu \beta}$, where $U^{\mu \alpha}_{\nu \beta}$ is the fourth order tensor whose elements are all equal to 1 and $I^{\mu \alpha}_{\nu \beta}$ is the delta tensor whose elements are equal to 1 if $\mu =\nu$ and $\alpha = \beta$, 0 otherwise. In the complete multiplex network without self-loops, a node in one level is connected with all its counterparts and all the other nodes in all levels except itself, so that a complete weighted undirected multiplex network consists of edges whose weights will always be understood to be $1$, except for self-loops whose weight is zero.

Degree and strength for a given node $i$ on layer $a$ will be denoted in general by $d_{i,a}$ and $s_{i,a}$, respectively, whereas the degree and strength of node $i$ with respect to the whole network by $d_i$ and $s_i$. More precisely, we define strength centrality matrix the $N\times L$ matrix whose entries are the strengths of each node in each level: $S^{\mu\alpha}=M^{\mu \alpha}_{\nu\beta}u^{\beta}u^{\nu}$, where $u^{\beta}$ and $u^{\nu}$ are the all $1$'s $L-$ and $N-$vectors, respectively. The total strength of a specific node $i$ on layer $a$ is then given by $s_{i,a}=S^{\nu\beta}E_{\nu\beta}(i,a)$, whereas the global strength of a specific node $i$ is then given by $s_{i}=\sum_{a=1}^{L}s_{i,a}$. We can finally define the total strength of a level as $s_{a}=\sum_{i=1}^{N}s_{i,a}$. Similarly for the degrees by replacing $M^{\mu \alpha}_{\nu\beta}$ by $A^{\mu \alpha}_{\nu\beta}$. Notice that a node such that $d_{i,a}=1$ is called a pendant node.

Throughout the text, we will adopt the Einstein’s summation convention: the summation symbol is omitted for sums over repeated indices. In particular, we will use tensors contraction, by setting in a tensor a couple of indices equal, in order to sum with respect to layers, nodes or both. This operation reduces the order of the tensor by 2.

\section{Triangles in multiplex networks}
\label{Triangles in multiplex networks}

In the literature, the definition of local clustering coefficient is related to the ratio between (weighted or unweighted) actual and potential triangles around a node \cite{Watts_1998}. Our aim is to propose coefficients for multiplex networks preserving the same meaning. To formally define the clustering coefficients, we need to define first what a triangle in a multiplex network is.
In a multiplex network in which non-diagonal couplings are allowed, a triangle is meant to be a closed triplet $i,j,k$ such that the three nodes can belong to up to three different levels and they are connected by inter or intra-layer links. By this definition, we mean to include all possible closed triplets, moving in all directions, along inter or intra-layer links. This definition extends the one adopted in \cite{DeDomenico2013} for a monoplex unweighted network

Since two possible orientations are associated to each undirected triangle, starting from and returning to the same node, we can consider each triangle equivalent to two 3-cycles. In particular, in a monoplex unweighted network without self-loops, the number of actual 3-cycles to which node $i$ belongs is given by $t(i)=A^\mu_\nu A^\nu_\rho A^\rho_\sigma e_{\mu}(i) e^{\sigma}(i)$, where $A^\mu_\nu$ is the binary adjacency matrix. On the same line, the number of triplets around $i$, i.e. of potential 3-cycles to which $i$ belongs, is given by the formula $t_p(i)=A^\mu_\nu F^\nu_\rho A^\rho_\sigma e_{\mu}(i) e^{\sigma}(i)$.

Keeping this in mind, in a multiplex unweighted network we define:

\begin{equation}
t(i,a)=A^{\mu \alpha}_{\nu \beta}A^{\nu \beta}_{\rho \gamma}A^{\rho \gamma}_{\sigma \delta}E_{\mu \alpha}(i,a)E^{\sigma \delta}(i,a)
\label{triangles_node_level}
\end{equation}

\noindent where $A^{\mu \alpha}_{\nu \beta}$ is the binary adjacency tensor.
Formula (\ref{triangles_node_level}) counts the number of actual 3-cycles to which node $i$ on level $a$ belongs, being the links in the triangles on the same level $a$ or not. 
Applying formula (\ref{triangles_node_level}), we can compute the total number of 3-cycles to which nodes belong on all levels or similarly, the total number of 3-cycles formed by all the nodes on a given level $a$. 

In other words, formula (\ref{triangles_node_level}) is flexible enough to allow us to choose in the multiplex network different scales of observations, simply by applying a suitable contraction over indices (by nodes or by layers).
Indeed, by contracting over all the levels on which node $i$ lies, we obtain the total number $t_{\rm N}(i)$ of 3-cycles to which $i$ belongs:

\begin{equation}\label{triangles_node}
t_{\rm N}(i)=A^{\mu \alpha}_{\nu \beta}A^{\nu \beta}_{\rho \gamma}A^{\rho \gamma}_{\sigma \alpha}E_{\mu}^{\sigma}(i)
\end{equation}

On the other hand, by contracting over all the nodes on the same layer $a$ we get the number $t_{\rm L}(a)$ of 3-cycles to which all nodes on level $a$ belong

\begin{equation}\label{triangles_level}
t_{\rm L}(a)=A^{\mu \alpha}_{\nu \beta}A^{\nu \beta}_{\rho \gamma}A^{\rho \gamma}_{\mu \delta}E_{\alpha}^{\delta}(a)
\end{equation}

Finally, by contracting over all nodes and layers, we obtain the total number of 3-cycles in the multiplex network:

\begin{equation}\label{triangles}
t=A^{\mu \alpha}_{\nu \beta}A^{\nu \beta}_{\rho \gamma}A^{\rho \gamma}_{\mu \alpha}
\end{equation}

\begin{figure}[H]
	\centering
	\includegraphics[scale=0.4]{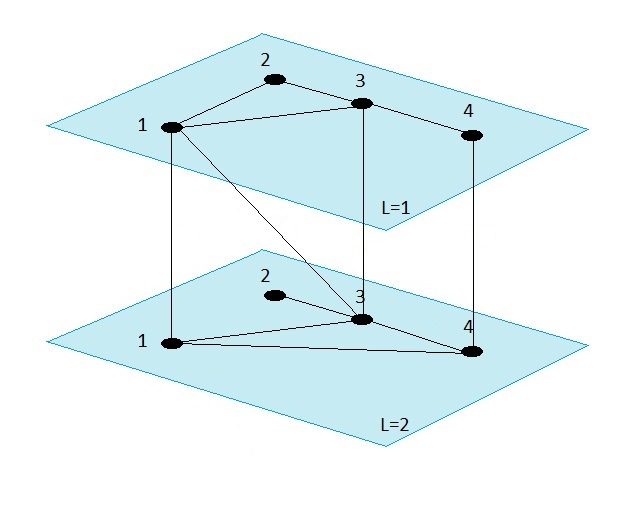}
	\caption{Example of a multiplex network with $N=4$ nodes and $L=2$ layers}
	\label{fig1}
\end{figure}

Let us consider, for example, the simple multiplex network with $N=4$ nodes and $L=2$ layers in Figure \ref{fig1}. We highlight the fact that this toy example exhibits a link between levels of a non-diagonal type, precisely between nodes $1$ and $3$ on the first and second layer. For node 1 on layer 1, for instance, we obtain $t(1,1)=6$; indeed, it belongs to one intra-layer triangle (with nodes 2 and 3 on layer 1) and two inter-layer triangles (with nodes 1 and 3 on layer 2 and node 3 on layers 1 and 2). Similarly, $t(1,2)=4$ (for the node 1 on layer 2 we count one triangle on layer 2 and one inter-layer triangle).
 
\noindent The total number of 3-cycles to which node 1 in all layers belongs to is then $t_{\rm N}(1)=10$ (2 intra-layer triangles and 3 inter-layer triangles). Notice that the inter-layer triangle with two vertices in node 1 in layer 1 and node 1 in layer 2 counts twice. This choice is consistent with the idea that the total $3-$cycles to which a node belongs must be counted and, although they are identifiable, the two nodes 1 on the two levels represent a different characterization of the same object. Focusing on each layer, we have $t_{\rm L}(1)=12$. Again, following the same argument, the triangle with vertices in nodes 1, 2, and 3 on layer 1 counts three times since we are counting the total number of $3-$cycles to which all nodes on such layer belong to. It can be easily checked that also $t_{\rm L}(2)=12$ so that, finally, $t=24$ for the whole multiplex network.

\section{Clustering coefficients in multiplex networks}
\label{Clustering for multiplex networks}

\subsection{General formulation of clustering coefficients}
\label{General definitions}

In this section we introduce the general definitions for local and global clustering coefficients on multiplex networks. Clustering coefficient is defined as the ratio between the number (or the weight) of actual triangles to which a node belongs, and the number (or the weight) of potential triangles to which it could belong. Of course, in an undirected network, this is equivalent to consider the ratio between the corresponding numbers of 3-cycles.\footnote{For a concise but comprehensive presentation of the clustering coefficients on monoplex networks, the reader can refer to section 1 in the Supplementary Material. In the present section, we deal directly with the extension, still not so much discussed in the literature, of those concepts to the case of multiplex networks as defined in the text.}

Following the line of the previous section, in a multiplex network, we have at first to decide where to take actual or potential triangles around a given node. Indeed, being a multiplex network a complex object, we are interested in representing the clustering structure from different scales of observation. This is the reason why, in this framework, we define four different types of clustering coefficients, namely three local coefficients, in dependence on which node and/or level is taken into account, and a global one, for the whole network. 

We start by providing a general expression for each of the four types of coefficients. The rationale behind each of the definitions is always to take the ratio between the number of real triangles and the number of potential triangles. The difference between the four types lies in the choice of where to take triangles: around a single node on a single level, around a node in all its levels, around all the nodes on a single level or around all nodes in all levels. The computation of the number of triangles therefore reflects the possibilities described in section \ref{Triangles in multiplex networks} and summarized by the formulas (\ref{triangles_node_level}-\ref{triangles}). From a mathematical point of view, these computations are performed by contracting three different tensors, in the most appropriate way, depending on the clustering structure we want to catch, in such a way to count the desired $3-$cycles. To do this, we initially propose very general expressions which will subsequently be declined in concrete and applied expressions of the same coefficients.

Let $H$ be any adjacency-like tensor with components $H^{\mu \alpha}_{\nu \beta}$. By adjacency-like tensor we mean a general representation of the adjacency relations in a multiplex network, standing for simple binary adjacency relations or weighted adjacency ones.  In particular, $H^{\mu \alpha}_{\nu \beta}$ will be replaced by $A^{\mu \alpha}_{\nu \beta}$ or (possibly normalised) $M^{\mu \alpha}_{\nu \beta}$ and, as we will see, the actual choice of $H^{\mu \alpha}_{\nu \beta}$ will depend on the specific definition of the clustering coefficient. In particular, we define:

\begin{enumerate}
	
	\item Clustering coefficient of node $i$ on level $a$:
	
	\begin{equation}\label{clust_nodo_livello}
		C(i,a)=\frac{[{H_1}]^{\mu \alpha}_{\nu \beta}[{H_2}]^{\nu \beta}_{\rho \gamma}[{H_3}]^{\rho \gamma}_{\sigma \delta}E_{\mu \alpha}(i,a)E^{\sigma \delta}(i,a)}{[{H_4}]^{\mu \alpha}_{\nu \beta}[F]^{\nu \beta}_{\rho \gamma}[{H_5}]^{\rho \gamma}_{\sigma \delta}E_{\mu \alpha}(i,a)E^{\sigma \delta}(i,a)}
	\end{equation}
	
	$C(i,a)$ represents the local coefficient\footnote{It is worth to stress that the number of potential triangles in the multiplex network can be obtained by suitable contractions of tensor $H$ with $F$, being the latter the adjacency tensor of the complete unweighted network introduced in section \ref{Mathematics of multiplex networks}.} for a node $i$ on a single layer $a$ and it takes into account all intra and inter-layer triangles that node forms with other pairs of nodes in the network, according to formula (\ref{triangles_node_level}). It gives a measure of how much that node in that level is clustered in the whole multiplex network.
	
	\item Clustering coefficient of node $i$ (over all the levels to which it belongs):
	
	\begin{equation}\label{clust_nodo}
		C_{\rm N}(i)=\frac{[H_{1}]^{\mu \alpha}_{\nu \beta}[H_{2}]^{\nu \beta}_{\rho \gamma}[H_{3}]^{\rho \gamma}_{\sigma \alpha}E_{\mu}^{\sigma}(i)}{[H_{4}]^{\mu \alpha}_{\nu \beta}[F]^{\nu \beta}_{\rho \gamma}[H_{5}]^{\rho \gamma}_{\sigma \alpha}E_{\mu}^{\sigma}(i)}
	\end{equation}
	
	$C_{\rm N}(i)$ is the coefficient for a node $i$ when the node itself is globally considered on all levels. More precisely, it takes into account all triangles to which $i$ belongs in all levels in the whole multiplex network according to formula (\ref{triangles_node}).
	
	\item Clustering coefficient of level $a$ (over all the nodes on the level):
	
	\begin{equation}\label{clust_livello}
		C_{\rm L}(a)=\frac{[H_{1}]^{\mu \alpha}_{\nu \beta}[H_{2}]^{\nu \beta}_{\rho \gamma}[H_{3}]^{\rho \gamma}_{\mu \delta}E_{\alpha}^{\delta}(a)}{[H_{4}]^{\mu \alpha}_{\nu \beta}[F]^{\nu \beta}_{\rho \gamma}[H_{5}]^{\rho \gamma}_{\mu \delta}E_{\alpha}^{\delta}(a)}
	\end{equation}

$C_{\rm L}(a)$ provides a global coefficient referring to a specific layer $a$. 
In this sense it gives an idea of how much clustered the whole layer is. It is worth noting that $C_{\rm L}(a)$ could be different from the global clustering coefficient of the layer, computed using the adjacency relations only between nodes belonging to the same layer. Indeed, it considers all types of triangles described by formula (\ref{triangles_level}), whether they are lying on the layer or not.
	
	\item Global clustering coefficient of the whole network:
	
	\begin{equation}\label{clust_total}
		C=\frac{[H_{1}]^{\mu \alpha}_{\nu \beta}[H_{2}]^{\nu \beta}_{\rho \gamma}[H_{3}]^{\rho \gamma}_{\mu \alpha}}{[H_{4}]^{\mu \alpha}_{\nu \beta}[F]^{\nu \beta}_{\rho \gamma}[H_{5}]^{\rho \gamma}_{\mu \alpha}}
	\end{equation}

This coefficient considers all triangles of all nodes in all levels in line with formula (\ref{triangles}). 	Note that this global coefficient represents a generalization of the well-known definition of transitivity present in the literature (see, for instance, \cite{Newman2010}) as the ratio between number of closed paths of length two and number of paths of length two. See section 1 in the Supplementary Material for further details.
\end{enumerate}

These definitions allow to extend with a general formula in a unified approach all the coefficients already existing in the literature for monoplex networks, by properly setting the adjacency tensors $[H_{k}]^{\mu \alpha}_{\nu \beta}$, as we will show in the next subsection.

\subsection{Relations with clustering coefficients for monoplex networks}
\label{Main clustering coefficients}

In a weighted multiplex network, both intra-layer links and inter-layer links are weighted and all tensors are symmetric.\footnote{Symmetries of $4^{\rm th}$ order tensors present a richer set of possibilities than the symmetry of $2^{\rm nd}$ order tensors, since a number of symmetries can be defined by applying different 'symmetry rules' on the four coefficient indices. Indeed, we may have major symmetry, minor symmetry and total symmetry. We refer here to the major symmetry whose rule is $H^{\mu \alpha}_{\nu \beta}=H_{\mu \alpha}^{\nu \beta}$}
Let us denote by

\begin{itemize}
	\item $M$ the weighted adjacency tensor;
	\item $A$ the corresponding binary adjacency tensor;
	\item $\tilde{M}=\frac{1}{\cal W}M$ the normalised adjacency tensor, where ${\cal W}=\max_{\mu \nu \alpha \beta}M^{\mu \alpha}_{\nu \beta}$;
	\item $\hat{M}= \tilde{M}^{1/3}$ the classical entry-wise cubic root of $\tilde{M}$.
\end{itemize}

For the sake of brevity, we refer here only to the local coefficient $C(i,a)$, defined by formula (\ref{clust_nodo_livello}). However, other clustering coefficients defined by formulas (\ref{clust_nodo}), (\ref{clust_livello}) and (\ref{clust_total}) can be adapted in a similar manner. For monoplex undirected networks the most important clustering coefficients in the literature are provided by \cite{DeDomenico2013, Barrat2004, Onnela2005}. Their definitions and properties are summarized in section 1 of the Supplementary Material. Our coefficients generalize them in the multiplex context as follows:

\begin{enumerate}
	\item Clustering Coefficient $\tilde{C}(i,a)$:
	
	Setting ${H_k}=\tilde{M}$, for each $k=1,\dots, 5$, we obtain
	
	\begin{equation}\label{DeDomenico}
		\tilde{C}(i,a)=\frac{\tilde{M}^{\mu \alpha}_{\nu \beta}\tilde{M}^{\nu \beta}_{\rho \gamma}\tilde{M}^{\rho \gamma}_{\sigma \delta}E_{\mu \alpha}(i,a)E^{\sigma \delta}(i,a)}{\tilde{M}^{\mu \alpha}_{\nu \beta}F^{\nu \beta}_{\rho \gamma}\tilde{M}^{\rho \gamma}_{\sigma \delta}E_{\mu \alpha}(i,a)E^{\sigma \delta}(i,a)}
	\end{equation}
	
	The numerator considers the weighted actual triangles to which the node $i$ belongs in the layer $a$, according to formula (\ref{triangles_node_level}). The denominator considers the triangles in a network for which the neighbours of $i$ is completely connected, identifying all the potential triangles around that node.\footnote{In the computation of weighted potential triangles, we close the open triplets by adding a new link having weight equal to 1, which is the maximum possible weight in $\tilde{M}$. This is in line with the coefficient introduced by (\ref{DeDomenico}) for monoplex networks}\\
	Let us observe that, in formula (\ref{DeDomenico}), the weight attributed to each triangle is obtained as the product of the weights of the three links, that is by the (normalised) product of the corresponding components in the tensor $M$. Hence, formula (\ref{DeDomenico}) generalizes the clustering coefficient introduced in \cite{DeDomenico2013}.

	\item Clustering Coefficient $C(i,a)$:
	
	Setting ${H_1}={H_4}=M$ and ${H_2}={H_3}={H_5}=A$, we have
	
	\begin{equation}\label{Barrat}
		C(i,a)=\frac{M^{\mu \alpha}_{\nu \beta}A^{\nu \beta}_{\rho \gamma}A^{\rho \gamma}_{\sigma \delta}E_{\mu \alpha}(i,a)E^{\sigma \delta}(i,a)}{M^{\mu \alpha}_{\nu \beta}F^{\nu \beta}_{\rho \gamma}A^{\rho \gamma}_{\sigma \delta}E_{\mu \alpha}(i,a)E^{\sigma \delta}(i,a)}
	\end{equation}

	In this case, the weight attributed to each triangle is the sum of the weights of the links between node $i$ and its neighbours. In other words, it considers only two of the three links involved in the closed triplet, namely those adjacent to node $i$. Indeed, the tensor product in the numerator sums up these two weights when the triangular path is travelled clockwise first and counter-clockwise after and it requires the existence of the third link, not adjacent to node $i$, otherwise it vanishes. Notice that, in this coefficient, the presence of the tensor $M$ in the denominator makes unnecessary to normalize the tensor $M$. 
	Hence, formula (\ref{Barrat}) extends to the multiplex case the clustering coefficient proposed in \cite{Barrat2004} for monoplex networks.

	\item Clustering Coefficient $\hat{C}(i,a)$:
	
	Setting ${H_1}={H_2}={H_3}=\hat{M}$ and ${H_4}={H_5}=A$, we obtain
	
	\begin{equation}\label{Onnela}
		\hat{C}(i,a)=\frac{\hat{M}^{\mu \alpha}_{\nu \beta}\hat{M}^{\nu \beta}_{\rho \gamma}\hat{M}^{\rho \gamma}_{\sigma \delta}E_{\mu \alpha}(i,a)E^{\sigma \delta}(i,a)}{A^{\mu \alpha}_{\nu \beta}F^{\nu \beta}_{\rho \gamma}A^{\rho \gamma}_{\sigma \delta}E_{\mu \alpha}(i,a)E^{\sigma \delta}(i,a)}
	\end{equation}
	
	In formula (\ref{Onnela}) weights are assigned to each triangle as the geometric mean of the weights of the three links in the triangle itself. Indeed, the numerator reports the geometric mean of the corresponding components of the normalised tensor $M$, 
	whereas the denominator reports the product of tensor $A$ that counts the number of potential triangles to which $i$ belongs, ignoring the weights. 
	This fact makes our coefficient the immediate generalization to multiplex networks of the coefficient proposed in \cite{Onnela2005} for monoplex network.
\end{enumerate}

\subsection{Clustering coefficients via supradjacency representation}
\label{Clustering coefficient via matrix representation}

It is worth providing here a rewriting of the same coefficients in terms of \textit{supradjacency matrix}. Indeed, we can use the well-known unfolding procedure, also called flattening or matricization, to represent the adjacency tensor as a block matrix. This matrix, with $L$ square blocks, each one of order $N$, is called \textit{supradjacency matrix}:

\begin{equation}
\mathbf{W}=
\begin{bmatrix}
\mathbf{W}_{1} & \mathbf{W}_{12} & \cdots & \mathbf{W}_{1L} \\
\mathbf{W}_{21} & \mathbf{W}_{2} & \cdots & \mathbf{W}_{2L} \\
\vdots & \vdots & \ddots & \vdots \\
\mathbf{W}_{L1} & \mathbf{W}_{L2} & \cdots & \mathbf{W}_{L}
\end{bmatrix}
\end{equation}

\noindent where the diagonal blocks represent the weighted adjacency matrix of each layer $\textbf{W}_{aa}$, $a=1,...,L$ (denoted by $\textbf{W}_{a}$ for short), whereas the out off diagonal blocks $\textbf{W}_{ab}$ represent the weighted adjacency relations between nodes on layers $a$ and nodes on layer $b$. We denote its unweighted version by $\mathbf{A}$.

First, it is noteworthy that the fourth order tensor generated by the tensor product $M^{\mu \alpha}_{\nu \beta}M^{\nu \beta}_{\rho \gamma}M^{\rho \gamma}_{\sigma \alpha}$ can be represented, in a natural way, by the $NL$-square block matrix $\mathbf{W}^3$. It is also straightforward to observe that the number of triangles $t(i,a)$ provided by formula (\ref{triangles_node_level}) is the $i$-diagonal entry of the $a$-diagonal block, namely $[(\mathbf{A}^3)_a]_{ii}$. Notice that the supradjacency matrix $\mathbf{F}$, corresponding to the adjacency tensor $F^{\nu \beta}_{\rho \gamma}$ of the complete multiplex network, is the $NL$-square matrix having $1$ in all positions but the diagonal entries, where we have $0$.

In terms of supradjacency matrices, the representations of the coefficient $C(i,a)$ in (\ref{clust_nodo_livello}) for the three different versions in (\ref{DeDomenico}), (\ref{Barrat}) and (\ref{Onnela}) are, respectively

\begin{equation}\label{DeDomenico_supra}
	\tilde{C}(i,a)=\frac{[(\mathbf{\tilde{W}}^3)_a]_{ii}}{[(\mathbf{\tilde{W}F\tilde{W}})_a]_{ii}}
\end{equation}

\begin{equation}\label{Barrat_supra}
C(i,a)=\frac{[(\mathbf{WA}^2)_a]_{ii}}{[(\mathbf{WFA})_a]_{ii}}
\end{equation}

\begin{equation}\label{Onnela_supra}
	\hat{C}(i,a)=\frac{[ ( \mathbf{\hat{W}}^3)_{a}]_{ii}}{[(\mathbf{AFA})_a]_{ii}}
\end{equation}

Let us focus, for instance, on formula (\ref{Barrat_supra}). $[(\mathbf{WA}^2)_a]_{ii}$ is the $i$-diagonal entry of the $a$-diagonal block of the matrix $\mathbf{WA}^2$, and similarly for $[(\mathbf{WFA})_a]_{ii}$. 
Observe that the numerator in (\ref{Barrat_supra}) counts the number of actual triangles to which the node $i$ on level $a$ belongs. These triangles, weighted with the average weight of the links on $i$, can lie on level $a$ or even be outside level $a$. 
Furthermore, we have $[(\mathbf{WFA})_a]_{ii}=s_{i,a}(d_{i,a}-1)$, where $d_{i,a}$ and $s_{i,a}$ are degree and strength of $i$ in the layer $a$ (see the proof in section 2 of the Supplementary Material). Formula (\ref{Barrat_supra}) is the natural extension of the classical representation, in matrix terms, of the local clustering coefficient. But in the same way, we can represent clustering coefficient of the node $i$ in the whole network (formula (\ref{clust_nodo})):

\begin{equation}
C_{N}(i)=\frac{\sum_{a=1}^{L}[(\mathbf{WA}^2)_a]_{ii}}{\sum_{a=1}^{L}[(\mathbf{WFA})_a]_{ii}}
\end{equation}

Similarly, the clustering coefficient of a level $a$ over all its nodes (formula (\ref{clust_livello})) is:

\begin{equation}\label{M_livello Barrat}
C_{L}(a)=\frac{\sum_{i=1}^{N}[(\mathbf{WA}^2)_a]_{ii}}{\sum_{i=1}^{N}[(\mathbf{WFA})_a]_{ii}}
\end{equation}


Finally, the global clustering coefficient in formula (\ref{clust_total}) is

\begin{equation}\label{M_Barrat}
C=\frac{\sum_{a=1}^{L}\sum_{i=1}^{N}[(\mathbf{WA}^2)_a]_{ii}}{\sum_{a=1}^{L}\sum_{i=1}^{N}[(\mathbf{WFA})_a]_{ii}}=\frac{{\rm Tr}(\mathbf{WA}^2)}{{\rm Tr}(\mathbf{WFA})}
\end{equation}

\section{An illustrative example}
\label{toyex}

To illustrate the meaning of the clustering coefficients discussed in the previous sections we provide a simple example based on a small network with 4 nodes and 2 layers. 

\begin{figure}[H]
	\centering
	\includegraphics[scale=0.4]{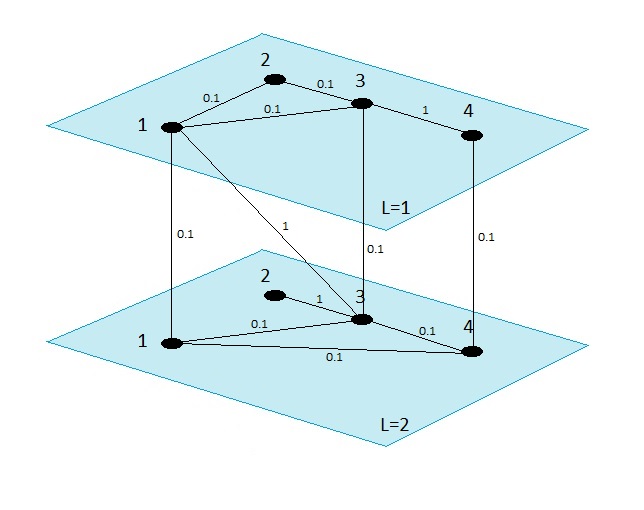}
	\caption{A simple weighted and undirected multiplex network, $L=2$, $N=4$}
	\label{example}
\end{figure}

The network is weighted: both inter- and intra-layers links have weights equal to either $0.1$ or $1$. We argue that the different coefficients we propose take into accounts in different ways both the effects of triangles and weights.

We collect the values of the local clustering coefficients $C(i,a)$ for each node in each level, according to the three different versions discussed in the previous sections and we reported these values in Table \ref{table1}. Notice that, being node $2$ on level $2$ a pendant node, its clustering coefficient cannot be computed.\footnote{It should be observed that a pendant node, like node 2 on layer 2, does not belong to any real or potential triangle and therefore, for this reason, its clustering coefficient cannot be calculated. Sometimes a conventional value of zero is assigned to it. When a node, although belonging to one or more potential triangles, does not belong to any real triangle, the clustering coefficient is zero. This is the case of node 4 on layer 1. However, it is appropriate to clearly distinguish the two cases.}
On the contrary, node 2 at layer 1 is totally clustered with its adjacent neighbours, although with low weights. Nodes $1$ and $3$ are well clustered in both layers, and all these aspects are well captured by the coefficient $C(i,a)$. On the contrary, the effect of low weights of triangles is better reflected in the other coefficients $\tilde{C}(i,a)$ and $\hat{C}(i,a)$.

\begin{table}[H]
	\small
	\centering{}
	\begin{tabular}{|c||c|c|c|c||c|c|c|c|}
		\hline \hline
		\multicolumn{9}{|c|}{\bf Local Clustering Coefficients $C(i,a)$}\tabularnewline
		\hline 
		\multirow[c]{2}{*}{\bf Version} & \multicolumn{4}{c||}{\bf Level 1} & \multicolumn{4}{c|}{\bf Level 2} \tabularnewline
		\cline{2-9}
		& 1 & 2 & 3 & 4  & 1 & 2 & 3 & 4  \tabularnewline
		\hline 
		\centering
		$\tilde{C}(i,a)$ & $0.064$ & $0.100$ & $0.033$ & $0$ & $0.367$ & $-$ & $0.013$ & $0.033$  \tabularnewline
		${C(i,a)}$  & $0.615$ & $1.000$ & $0.103$ & $0$ & $0.667$ & $-$ & $0.261$ & $0.333$
		\tabularnewline
		$\hat{C}(i,a)$  & $0.088$ & $0.100$ & $0.053$ & $0$ & $0.105$ & $-$ & $0.053$ & $0.033$
		\tabularnewline
		\hline \hline
	\end{tabular}
	\caption{Local Clustering Coefficients for each node in each level}
	\label{table1}
\end{table}

The different behaviour becomes even clearer if we look at the rankings of the nodes produced by the different coefficients. $\tilde{C}(i,a)$ and $\hat{C}(i,a)$ rankings are almost completely overlapping, except on node $3$ at layer $2$, whereas $C(i,a)$ produces a different ranking pattern for the nodes. In particular, $C(i,a)$ coefficient assigns the maximum value to node 2 at layer 1 since it belongs to only one real and potential triangle, while $\tilde{C}(i,a)$ and $\hat{C}(i,a)$ are more affected by the weights of the edges in the same triangle.

\begin{table}[H]
	\small
	\centering{}
	\begin{tabular}{|c||c|c|c|c||c|c||c|}
		\hline \hline
		\multicolumn{8}{|c|}{\bf  Clustering Coefficients}\tabularnewline
		\hline 
		\multirow[c]{2}{*}{\bf Version} & \multicolumn{4}{c||}{$C_{\rm N}(i)$} & \multicolumn{2}{c||}{$C_{\rm L}(a)$} & \multirow[t]{2}{*}{$C$}\tabularnewline
		\cline{2-8}
		& 1 & 2 & 3 & 4 & Level 1 & Level 2 & Network\tabularnewline
		\hline 
		\centering
		$\tilde{C}$ & $0.089$ & $0.100$ & $0.016$ & $0.008$ & $0.043$ & $0.020$ & $0.027$ \tabularnewline
		$C$  & $0.622$ & $1.000$ & $0.214$ & $0.118$ & $0.330$  & $0.288$ & $0.308$ \tabularnewline
		$\hat{C}$ & $0.094$ & $0.100$ & $0.053$ & $0.025$ & $0.068$ & $0.059$ & $0.063$ \tabularnewline
		
		\hline \hline
	\end{tabular}
	\caption{Clustering Coefficients for each node $i$, for each level $a$ and for the whole network}
	\label{table3}
\end{table}

\begin{figure}[H]
	\centering
	\includegraphics[scale=0.4]{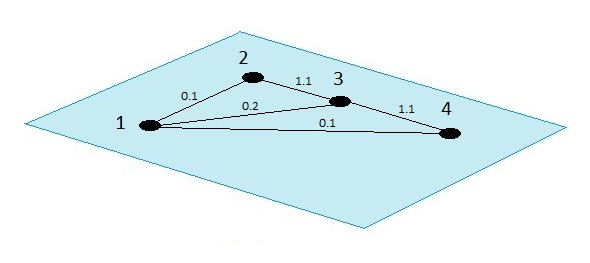}
	\caption{Overlay Network}
	\label{overlay}
\end{figure}

It is worth looking at the behaviour of the global coefficients (by nodes, and by layers for the whole network). Table \ref{table3} reports the global clustering coefficients for each node $C_{\rm N}(i)$, $i=1, 2, 3, 4$, for each level $C_{\rm L}(a)$, $a=1, 2$ and for the whole network $C$. On one hand, the three coefficients offer different interpretations of the nodes position in the whole network, through $C_{\rm N}$, in dependence of the number of actual triangles they belong to in each level. On the other hand,  $C_{\rm L}$ provides a view of the clustered structure level by level. In both cases, the proposed approach offers a more detailed picture of the structure than those that can be obtained by the global coefficient or by considering the overlay network. Table 3 reports the clustering coefficients for the overlay network, represented in figure \ref{overlay}.

\begin{table}[H]
	\small
	\centering{}
	\begin{tabular}{|c||c|c|c|c||c|}
		\hline \hline
		\multicolumn{6}{|c|}{\bf Clustering Coefficients (overlay network)}\tabularnewline
		\hline 
		\multirow[c]{2}{*}{\bf Version} & \multicolumn{4}{c||}{$C(i)$}  & {$C$}\tabularnewline
		\cline{2-6}
		& 1 & 2 & 3 & 4 &  Network\tabularnewline
		\hline 
		\centering
		$\tilde{C}_O$ & $0.800$ & $0.181$ & $0.024$ & $0.181$ & $0.063$  \tabularnewline
		$C_O$  & $0.170$ & $0.255$ & $0.170$ & $0.255$ & $0.191$  \tabularnewline
		$\hat{C}_O$  & $0.750$ & $1.000$ & $0.542$ & $1.000$ & $0.700$  \tabularnewline
		\hline \hline
	\end{tabular}
	\label{tableOverlay}
	\caption{Clustering Coefficients for the overlay network}
\end{table}

\section{Empirical application}
\label{empar}
\subsection{Data description and network construction}
In this section, we perform some empirical studies in order to assess the effectiveness of the proposed approaches. To this aim, we collected daily returns of a dataset referred to the time-period ranging from January 2001 to June 2017, that includes $102$ leading U.S. stocks constituents of the $S\&P$ 100 index at June 2017. Data have been downloaded from Bloomberg. Returns have been split by using monthly windows. The typical approach proposed in the literature about complex networks (see, e.g., \cite{Onnela2003b}, \cite{pozzi13}, \cite{Mantegna1999}) is to study this kind of data using time-varying networks. In other words, the common methodology is to build in each period (for instance, a month) a correlation network. \\ Hence, for each window, we have a network $G_{t}=(V_{t},E_{t})$ (with $t=1,...,T$, where $T=198$ for our dataset), where assets are nodes and links are weighted considering the correlation matrix $\boldsymbol{\Rho}_{t}=[_{t}\rho_{i,j}]_{i,j\in V_{t}}$, where  $_{t}\rho_{i,j}$ is the correlation coefficient between the empirical returns of a couple of assets $i$ and $j$ at time $t$ (with $i \neq j$). In order to assure that the weights range in the interval $[0,1]$, 
a meaningful solution has been proposed in \cite{Mantegna1999} (and adapted in \cite{Giudici2020}) based on distances $_{t}d_{i,j}$:
$_{t}d_{i,j}=1-\frac{1}{2}\sqrt{2(1-{}_{t}\rho_{i,j})}$.  The distance matrix
$\mathbf{D}_{t}=[{}_{t}d_{i,j}]_{i,j\in V_{t}}$,
with elements $0\leq{}_{t}d_{i,j}\leq1$, is then used as the weighted adjacency matrix of the graph $G_{t}$. \\

We propose here a different approach based on a multiplex network in order to catch also dependencies between assets that are observed over time. This specific structure also  allows to include in the analysis the possible presence of autocorrelation between stock returns over time (see, e.g., \cite{English}, \cite{Cont2007}). \\
 To this end, we compute the correlation matrix $\mathbf{\Rho}_{t}$ described before and the inter-temporal correlation matrices $\mathbf{\Rho}_{t,t+1}=[_{t,t+1}\rho_{i,j}]$  where $_{t,t+1}\rho_{i,j}$ is the correlation between the returns of asset $i$ at time $t$ and the returns of asset $j$ at time $t+1$. The coefficient $_{t,t+1}\rho_{i,j}$ is also computed for the case $i=j$ and it then represents the correlation between returns of the same asset in two subsequent time periods. As described above, it is possible to transform the correlation matrices in the distance matrices $\mathbf{D}_{t}$ and $\mathbf{D}_{t,t+1}$ and, hence, we 
construct a supradjacency matrix $\mathbf{W}$ 
of order $\left(N \cdot T\right)^{2}$. This matrix can be partitioned into $T^{2}$ blocks, which are represented by $N$-square matrices of order $N$. The blocks that belong to the main diagonal are represented by the matrices $\mathbf{D}_{t}$, while the off-diagonal blocks consider distances between couple of times $\mathbf{D}_{t, \tau}$ with $t=1,...,T$, $\tau=1,...,T$ and $\tau \neq t$).
It is noteworthy that we considered only correlations between subsequent time intervals (i.e. $\mathbf{D}_{t,t+1}$). Hence the elements of the matrices $\mathbf{D}_{t,\tau}$ with $\tau \neq t+1$ are equal to zero. 
This is motivated by the fact that we want to highlight, via inter-layer connections, the effects of dependencies between returns of subsequent periods. Obviously, this choice has the advantage of simplifying the structure of the multiplex network reducing the computational times.
	
\begin{figure}[!htb]
	\centering
	\includegraphics[width=1.0\textwidth]{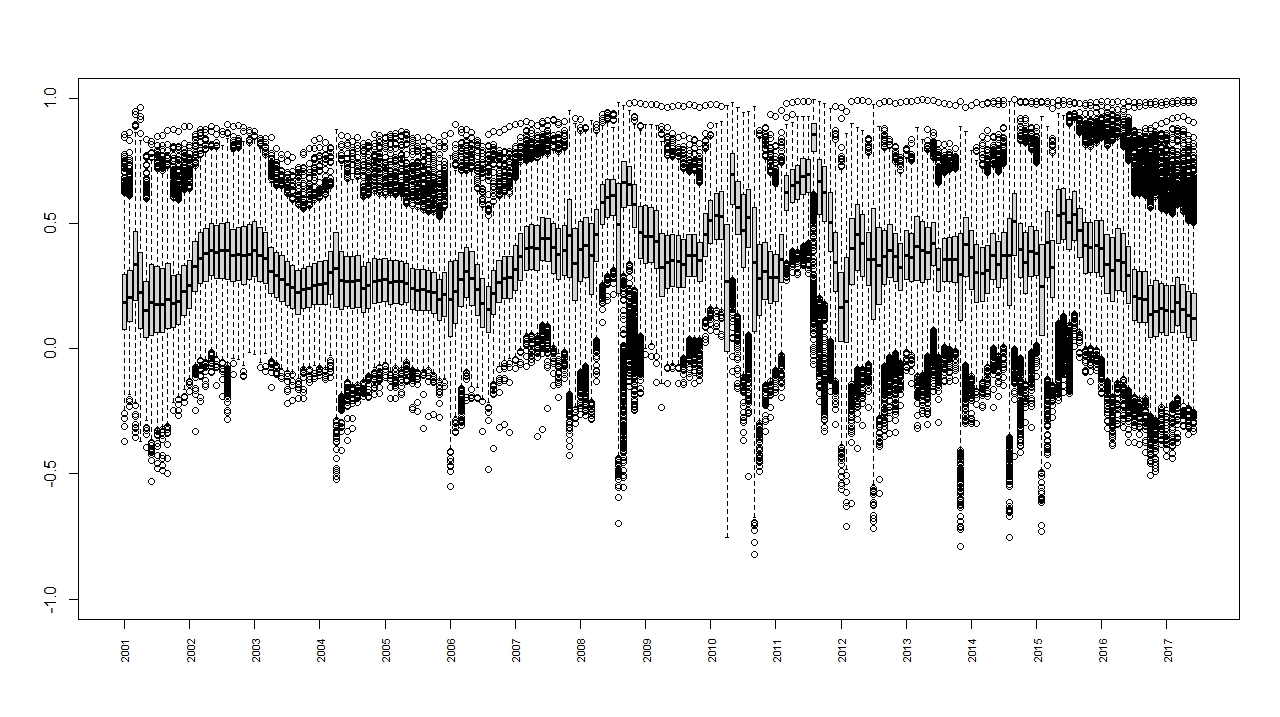}
	\caption{Distributions of elements of the correlation matrix $\boldsymbol{\Rho}_{t}$ for each time period. 
	}
	\label{fig:Cor}
\end{figure}

\begin{figure}[!htb]
	\centering
	\includegraphics[width=1.0\textwidth]{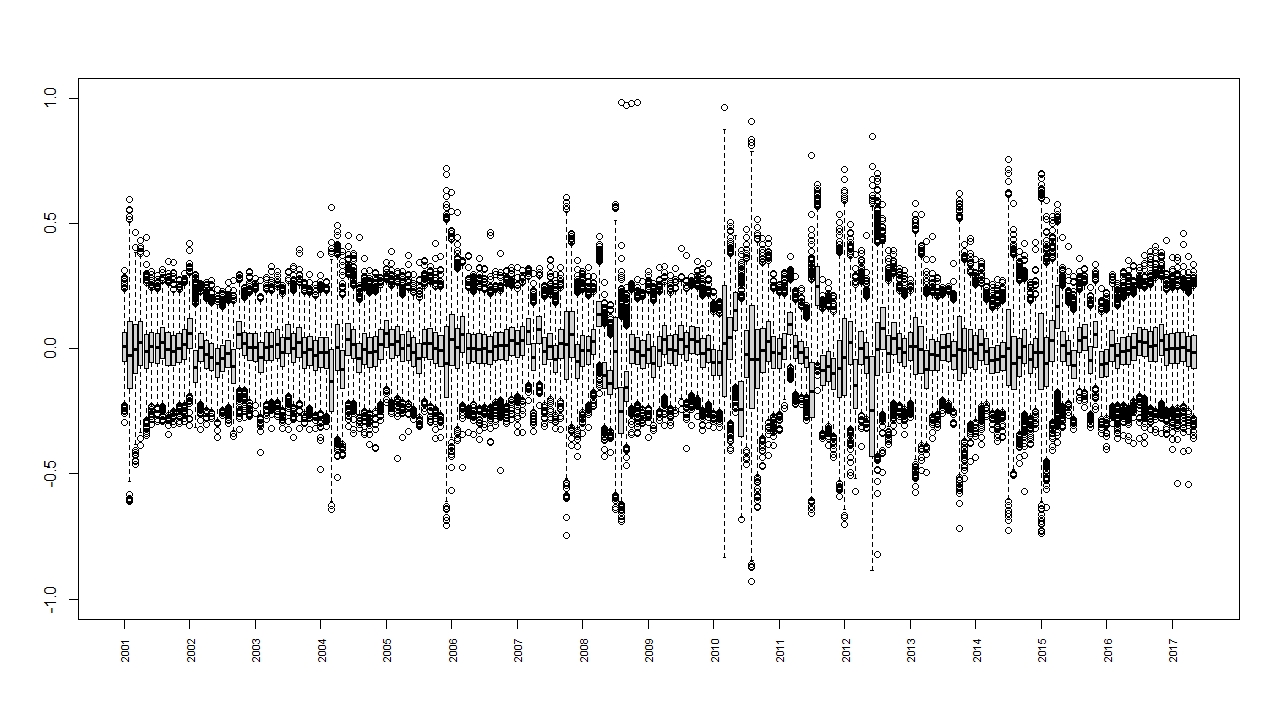}
	\caption{Distributions of elements of the correlation matrix $\boldsymbol{\Rho}_{t,t+1}$ for each couple of subsequent months. 
	}
	\label{fig:IntCor}
\end{figure}

Therefore, we build a multiplex network, where each layer considers the correlation between assets at time $t$ while inter-layer links consider the correlation over time. 
To give a first representation of the data, we report in Figures \ref{fig:Cor} and \ref{fig:IntCor} the distributions of the elements of the matrices $\boldsymbol{\Rho}_{t}$ and $\boldsymbol{\Rho}_{t,t+1}$. Values depend on assets' correlations in the same time period $t$ and in two subsequent periods ($t$, $t+1$), respectively. 

In Figure \ref{fig:Cor}, we observe on average a positive correlation between assets but, at the same time, a large number of assets that allow for diversification being negative correlated. Results confirm the compelling empirical evidence that the correlation structure among returns of the assets cannot be assumed to be constant over time (see, e.g., \cite{Forbes}, \cite{Bhattacharyya2018}, \cite{Wied}). In particular, in periods  of  financial  crisis,  correlations  among  stock  returns  increase,  a  phenomenon which is sometimes referred to as diversification meltdown (see, for instance, 2007-2008 and 2010-2011). \\
It is also noteworthy that on average we have a lower level of correlation over time (see Figure \ref{fig:IntCor}). This result appears in line with papers in the literature (see, e.g. \cite{Cont2007}).  However, despite this average pattern, a larger variability between assets is observed as well as some specific behaviours in specific time periods (see, for instance, 2007 and 2011). 

\begin{figure}[!htb]
	\centering
	\includegraphics[width=0.7\textwidth]{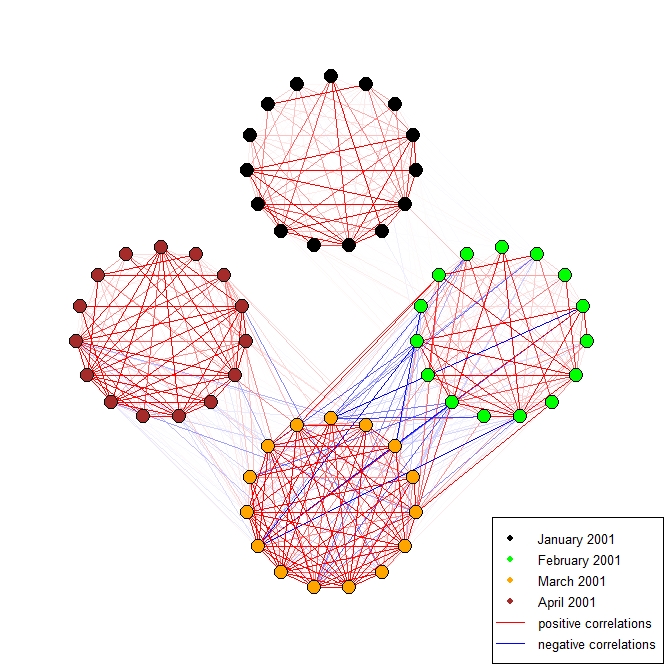}
	\caption{Excerpt of multiplex network. We report only fifteen assets and four layers (i.e. related to the period January-April 2001). Only for plotting purposes, we distinguish between positive and negative correlations. 
	}
	\label{fig:FigureNet}
\end{figure}

To provide a visual inspection of the complexity and the structure of the network, we display in Figure	\ref{fig:FigureNet} an excerpt of the whole multiplex network. In particular, for the sake of simplicity, first fifteen assets and four layers are reported. Furthermore, we plot edge weights using the linear correlation in order to emphasize positive and negative coefficients. However, as described above, the edge weights are then obtained by means of the distances to assure values in the interval $[0,1]$.
It could be observed a prevalence of positive correlations in each layer as well as the presence of negative coefficients over time (see for instance the connections between layer 2, February 2001, and layer 3, March 2001).

\subsection{Main Results: a comparison between alternative clustering coefficients}

\begin{figure}[!htb]
	\centering
	\includegraphics[width=6.5cm, height=5cm]{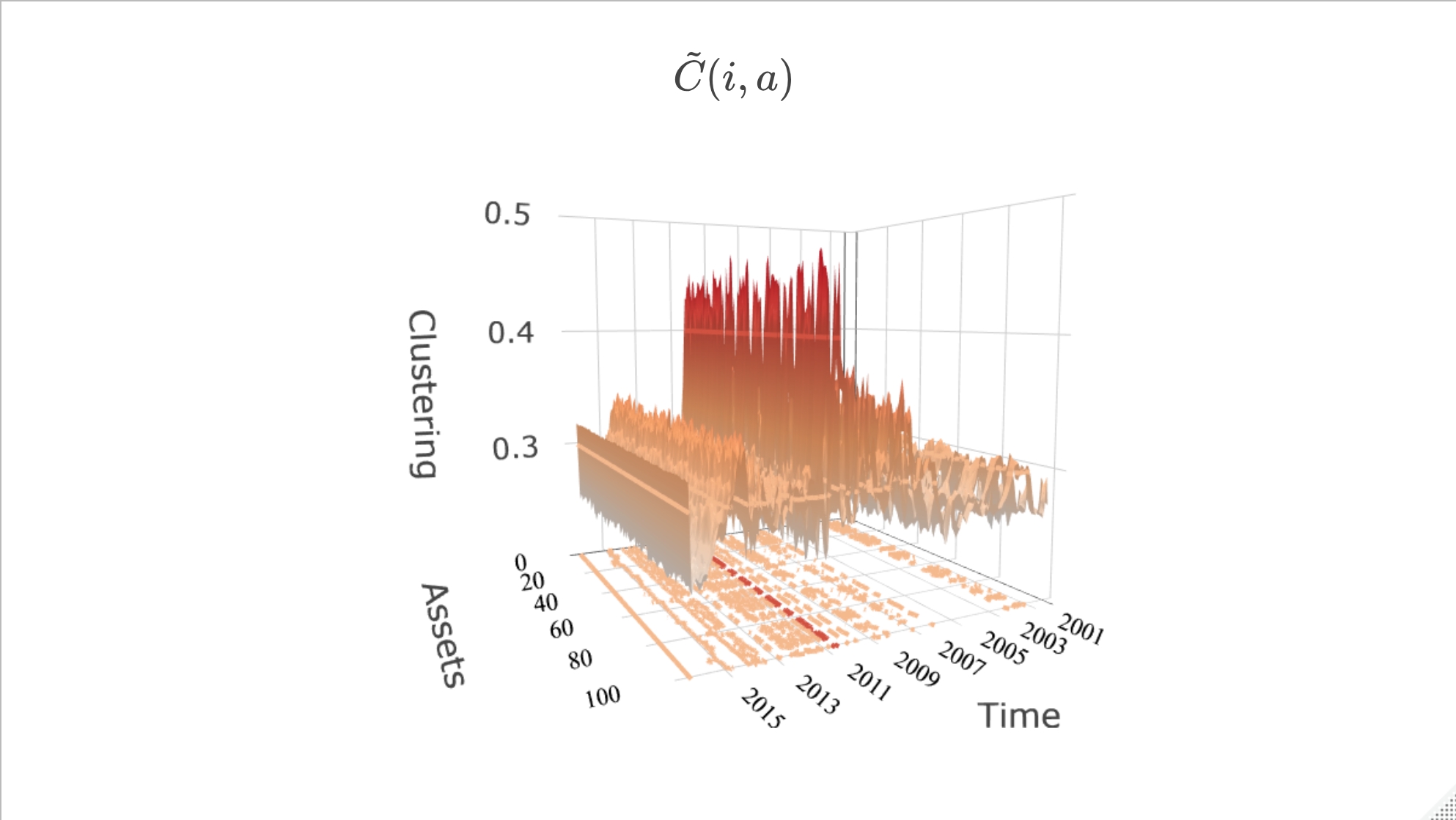}
	\includegraphics[width=6.5cm, height=5cm]{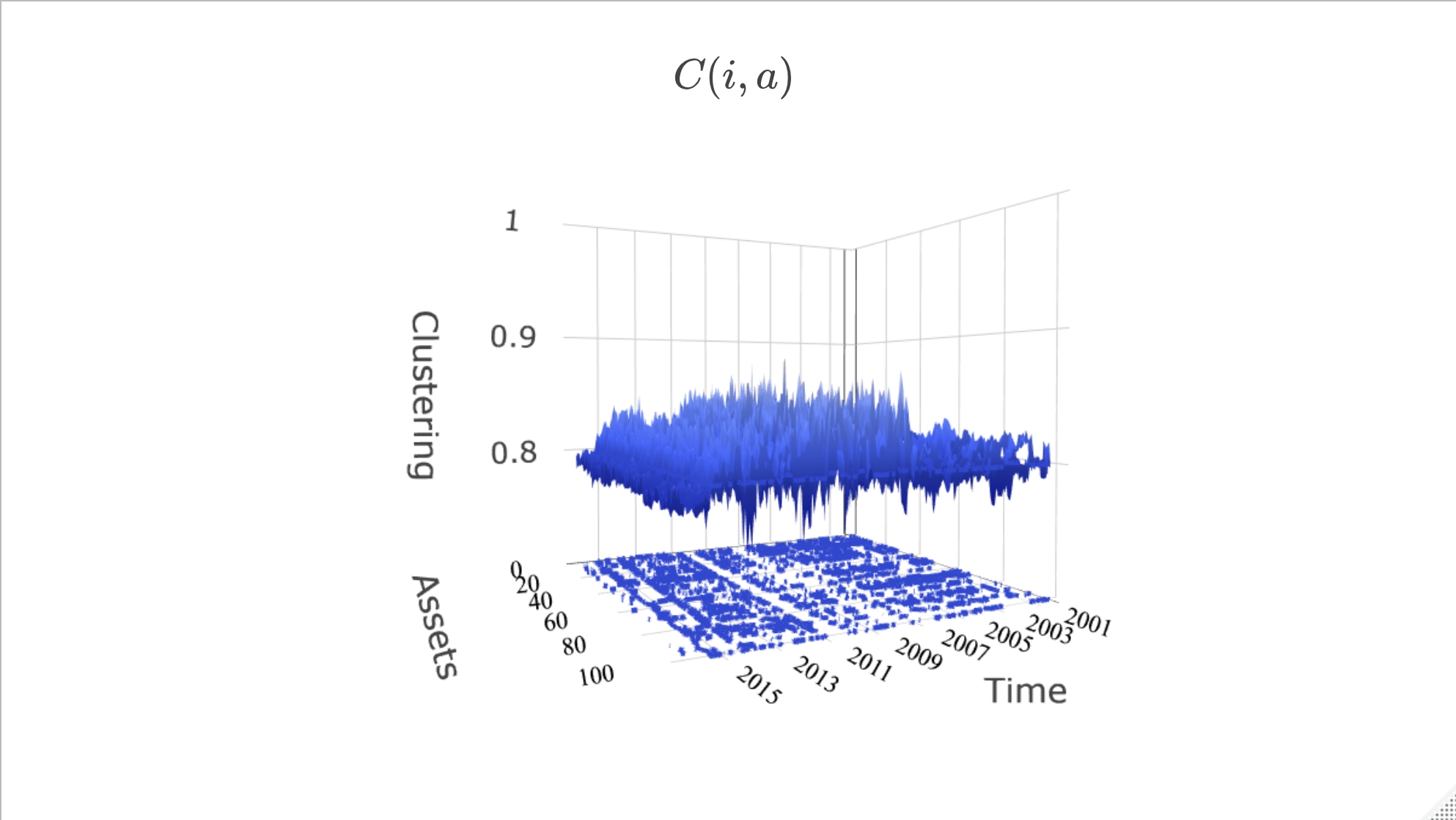}
	\includegraphics[width=6.5cm, height=5cm]{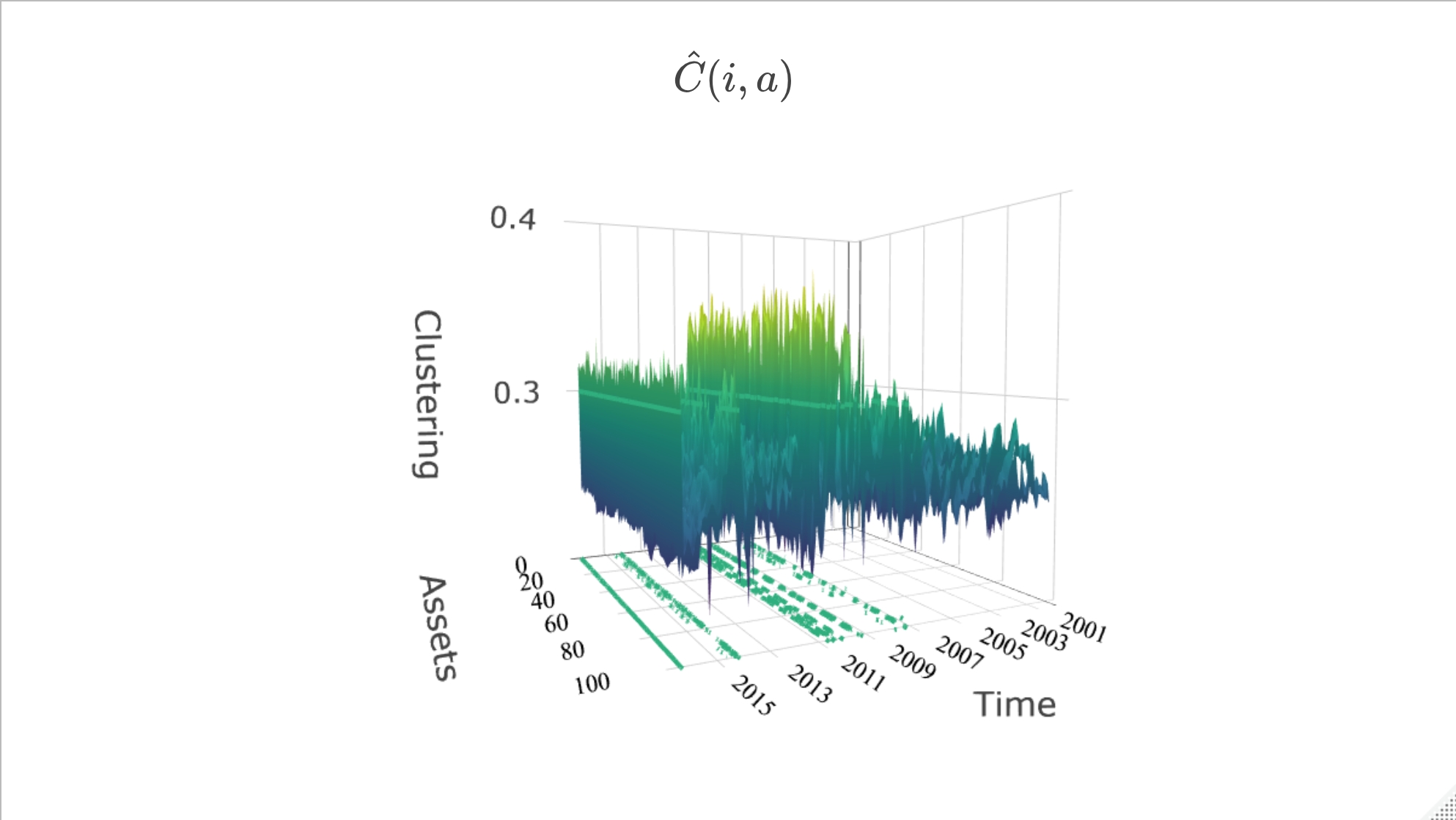}
	
	\caption{Local clustering coefficients for each node and layer based on formulas (\ref{DeDomenico_supra}), (\ref{Barrat_supra}) and (\ref{Onnela_supra})
	}
	\label{fig:FigureCompar}
\end{figure}

Alternative coefficients described in section \ref{Clustering coefficient via matrix representation} have been tested on the whole multiplex network. 
We start by comparing the local coefficients provided in formulas (\ref{DeDomenico_supra}), (\ref{Barrat_supra}) and (\ref{Onnela_supra}). Specific patterns with respect to each asset and time period are reported in Figure \ref{fig:FigureCompar}. \\
It is noteworthy how the clustering coefficients evolve consistently with the underlying financial events. This result is in line with the literature on monoplex networks (see, e.g., \cite{Minoiu}, \cite{Tabak2014}, \cite{CleGra}, \cite{Peder}). We observe that clustering coefficients tend to be lower in quiet periods and rise during crises. Sharper spikes occur around highly popular events that caused severe stress in the global financial system. The 2008 episode stands out as a large perturbation to the network, with higher coefficients in the second half of 2008 (the period that covers the Lehman Brothers failure and the rescue from bankruptcy of AIG). Then, a decline is observed until 2010, when a greater focus emerged on sovereign debt in the Eurozone. A second peak is observed at the end of 2011 with the highest levels of clustering. This is also reflected in the higher volatility of clustering distribution, which is justified by the different behaviour of the assets in the sample. \\
Comparing the three coefficients, it is noticeable that coefficients $\tilde{C}(i,a)$ and $\hat{C}(i,a)$  provide similar average levels of interconnection, while $C(i,a)$, given by formula (\ref{Barrat_supra}), shows values that are closer to one. These differences can be easily explained looking at the structure of the coefficients.
$\tilde{C}(i,a)$ and $\hat{C}(i,a)$ are strongly affected by the weights’ values involved in the observed triangles. 
In particular, the low average and the high skewness of weight link distribution lead to lower local clustering coefficients. On the other hand, $C(i,a)$ tend to be more affected by the number of triangles than by the weights. Therefore, since monoplex networks on each layer are almost complete, higher clustering coefficients are obtained in this case.

\begin{figure}[!htb]
	\centering
	\includegraphics[width=12cm, height=6cm]{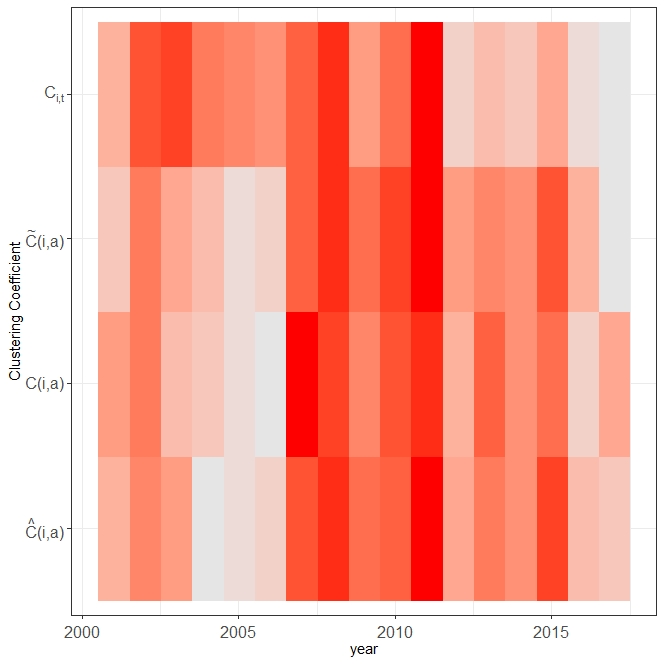}
	\caption{Ranking of average clustering coefficients for each year. The ranking has been computed separately for the coefficients $\hat{C}(i,a)$,  $C(i,a)$, $\tilde{C}(i,a)$ and $C_{i,t}$, averaged with respect to nodes and time periods. With $C_{i,t}$ we denote the coefficient provided in \cite{Onnela2005} applied separately to each monoplex network $G_{t}$. Darker bars indicate years characterized by higher clustering coefficients.
	}
	\label{fig:FigureClust}
\end{figure}

To better emphasize the differences between the alternative clustering coefficients, we display in Figure 
\ref{fig:FigureClust} the year-based ranking computed on the three coefficients $\hat{C}(i,a)$ $C(i,a)$, and $\tilde{C}(i,a)$. To obtain these rankings, we consider the local coefficients of the single node $i$ and layer $a$ (for instance, $\hat{C}(i,a)$) and we average them with respect to all the nodes and to the layers that refer to that year (for instance $a=1,...,12$ for the year 2001). The comparison is also extended to the application of the local clustering coefficient $C_{i,t}$ provided in \cite{Onnela2005} to each monoplex network $G_{t}$ with weights given by the matrices $\boldsymbol{D}_{t}$. In this case, correlations between two different periods are not considered. \\
Focusing on Figure \ref{fig:FigureClust}, a darker bar is associated to a higher clustering coefficient in that year.  We notice that, although different values are observed at local level, the three multiplex coefficients provide a very similar behaviour. Ranking based on $\hat{C}(i,a)$ and $\tilde{C}(i,a)$ show indeed a correlation of 0.95, easily justified by the high similarity of the two formulations. 
The coefficients $C(i,a)$, characterized by higher values in Figure \ref{fig:FigureCompar}, have however a closer behaviour in terms of ranking with a correlation around 0.85-0.90 with the other two coefficients. \\
Additionally, it is noticeable that both the multiplex and the monolayer solutions lead to the highest clustering coefficients in 2008 and 2011. Main differences between $C_{i,t}$ and the other coefficients are observed in the other periods. We have indeed that intermediate levels of interconnections are observed before than 2007 when the monolayer coefficient is adopted, while the multiplex versions provide a higher ranking to periods after 2011. 
Indeed, it is interesting to note that the inclusion of inter-layer dependencies provides differences in the ranking. The coefficient based on monoplex networks shows indeed a rank correlation, with the results based on multiplex formulas, that falls in the range $\left(0.4,0.5\right)$.

\subsection{An analysis at sectors levels}
\begin{figure}[!htb]
	\centering
	\includegraphics[width=6.5cm, height=5cm]{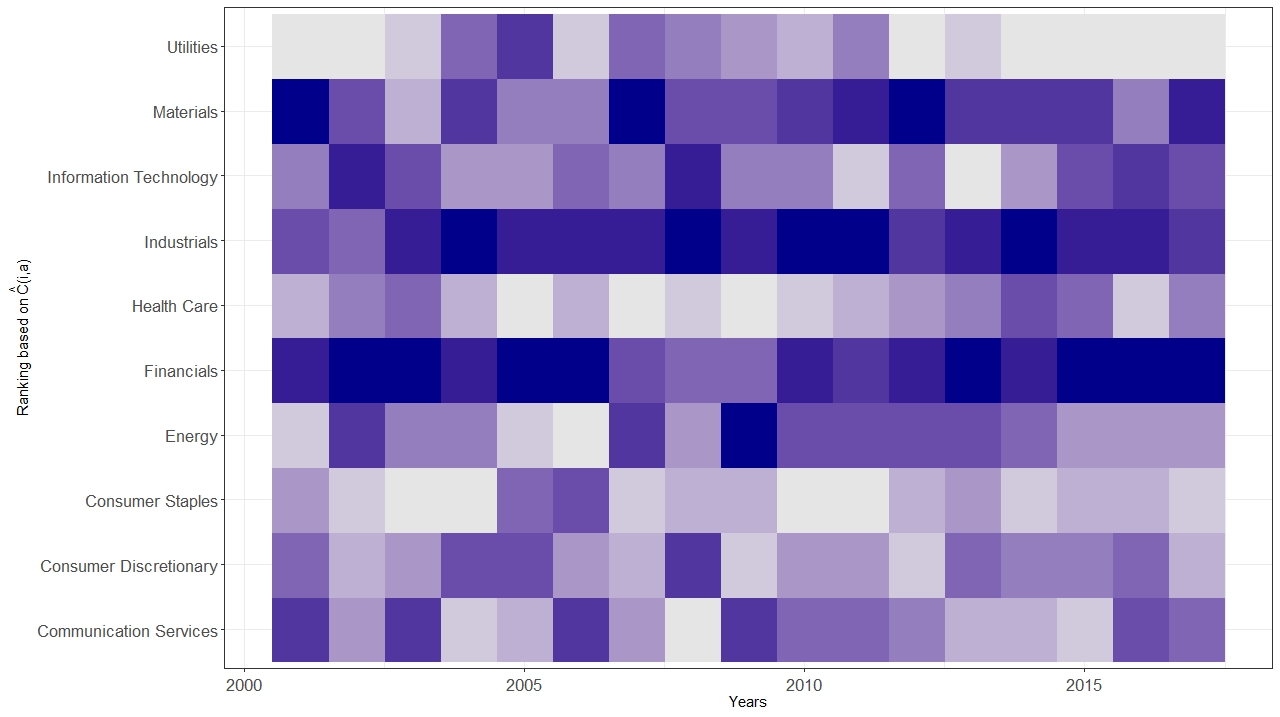}
	\includegraphics[width=6.5cm, height=5cm]{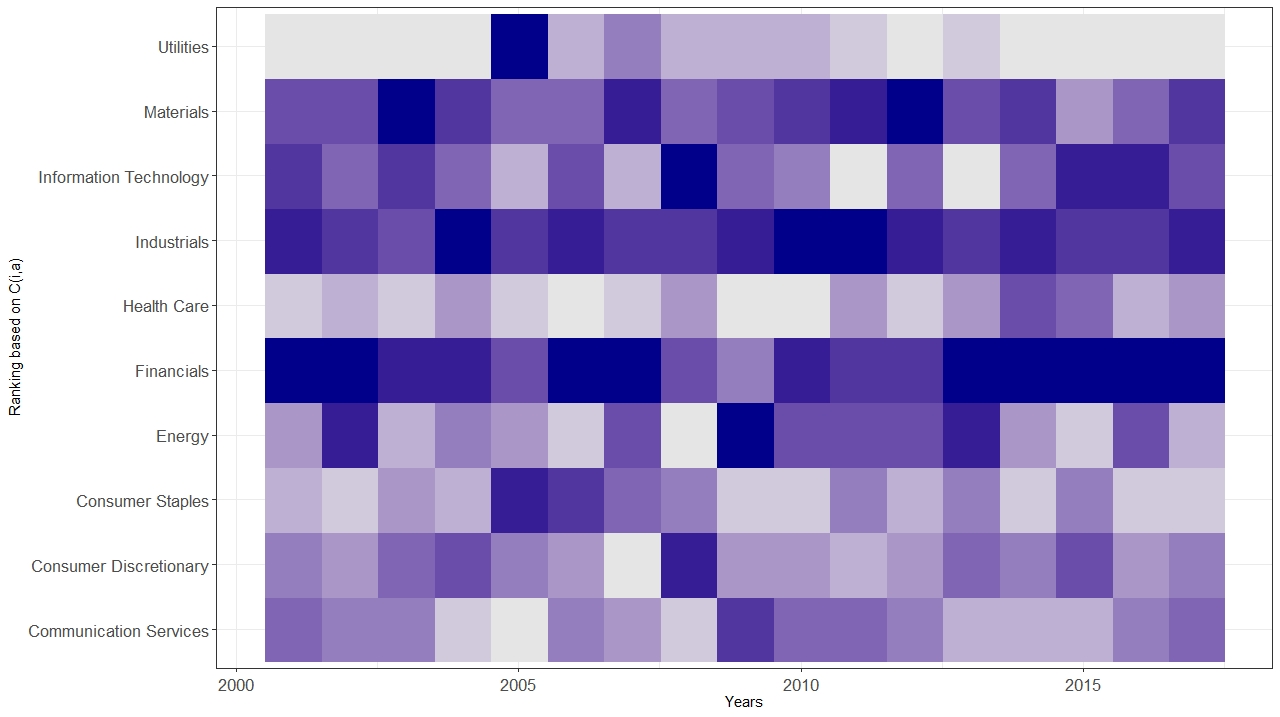}
	\includegraphics[width=6.5cm, height=5cm]{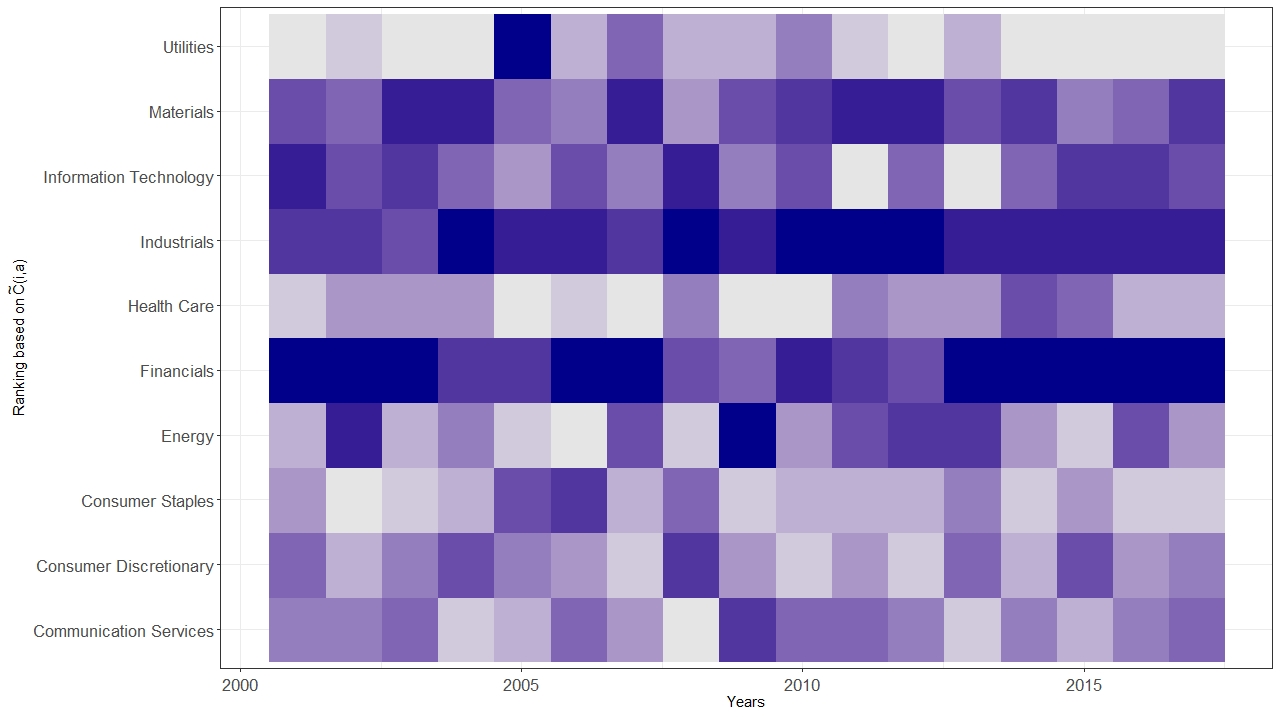}
	\includegraphics[width=6.5cm, height=5cm]{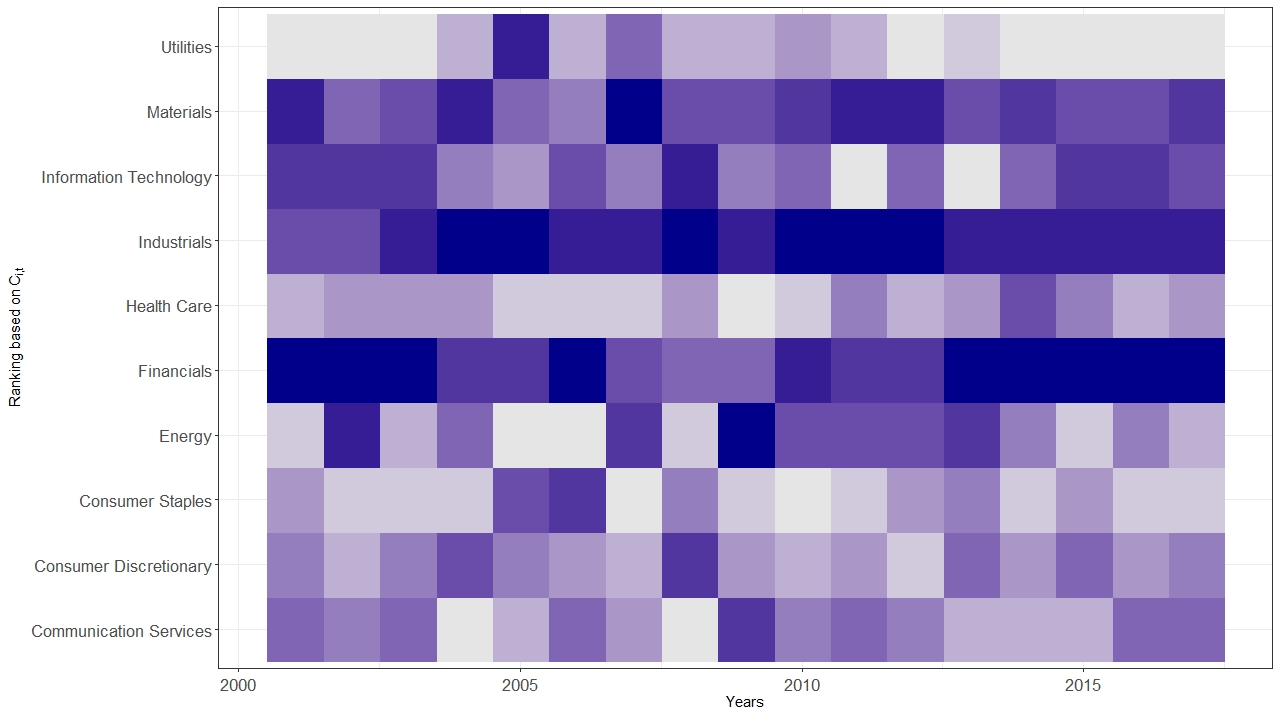}
	
	\caption{Ranking of local clustering coefficients of each sector. In clockwise order, plots are based respectively on coefficients $\hat{C}(i,a)$,  $C(i,a)$, $\tilde{C}(i,a)$ and $C_{i,t}$. Darker blue bars indicate sectors with higher clustering coefficients in that year.}
	\label{fig:Sectors}
\end{figure}

Attention has been also paid to the behaviour of assets that belong to the same sector. In particular, assets have been classified in ten sectors, according to the
standard sector classification defined by the Global Industry Classification Standard (GICS) developed by Morgan Stanley Capital International and Standard \& Poor’s.  Unlike other existing industry classifications (the Standard Industrial Classification System and the North American Industry Classification System), the GICS is based on the company’s principal business activity (i.e. its major sources of revenues or earnings). \\
The ten sectors are the following: communication services (TC), consumer discretionary (CD), consumer staple (CS), energy (EN), financials (FI), health care (HC), industrials (IN), information technology (IT), materials (MA) and utilities (UT). For a detailed description of sectors see, for instance, Appendix 1 in \cite{Beber}. \\ Therefore, we compute the clustering coefficient for a specific time period of each group as the simple average of the clustering coefficient of the individual nodes belonging to a given cluster (for a similar approach on monoplex networks, see, for instance, \cite{OnnelaJukka}). Sectors are then ordered in each year in a decreasing order on the basis of the clustering coefficients. The procedure has been applied separately considering the three different versions of multiplex coefficients and the monolayer coefficient provided in \cite{Onnela2005}. Main results are reported in Figure \ref{fig:Sectors}. \\

The analysis of sector dispersion and correlation is indeed a crucial point in financial markets. For instance, over the years, sector rotation investment strategies have been able to take advantage of environments with wider dispersions and lower correlations (see, e.g., \cite{Solnik}). As well known, investing in asset classes that demonstrate little or no correlation to one another may help to enhance diversification and reduce portfolio volatility. To this end, by means of the clustering coefficient, we catch in Figure \ref{fig:Sectors} which sectors are characterized by a higher level of interconnection in each time period. It is noticeable that all coefficients show on average that the UT sector is least related to the other sectors, while the FI and IN sectors are most related to the other sectors across all sample periods. Therefore, the results suggest that the UT sector has relatively weak dependence with the other sectors and offers relatively great diversification benefits. This is in line with the results found in \cite{Sukcharoen} on a different time period.
High relevance of IN and FI sectors can be explained by the fact that the IN sector is also fundamentally linked with the FI sector as companies in the IN sectors typically finance their capital projects through a financial institution. This finding is consistent with the result in \cite{Kim}.

\begin{figure}[!htb]
	\centering
	\includegraphics[width=15cm, height=8cm]{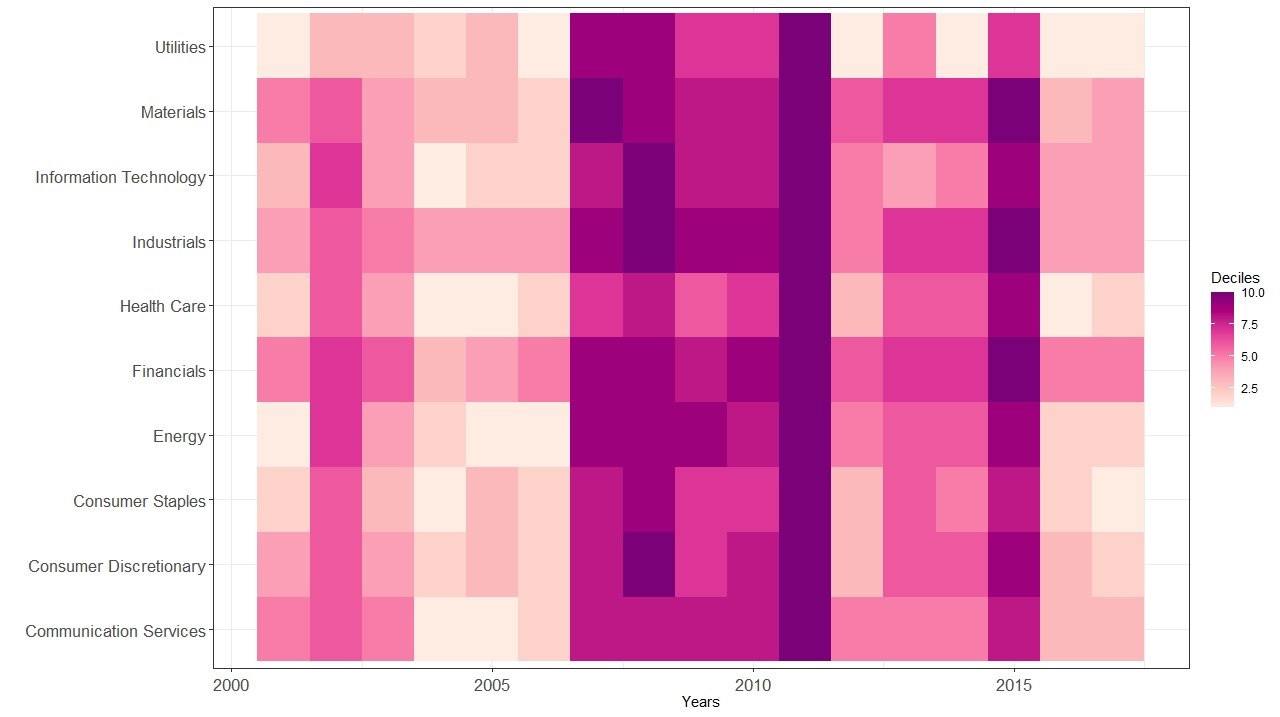}
	\caption{The whole distribution of average clustering coefficient of each sector (based on $\hat{C}(i,a)$) has been divided into ten deciles. The colour indicates the decile of the distribution to which the clustering belongs. In particular, darker bars indicate sectors with higher clustering coefficients in that year with respect to the whole period.}
	\label{fig:SectorsClust}
\end{figure}

While previous comments were based on ranking sectors separately for each calendar year, we focus now on the distribution of the average clustering coefficients for each sector considering the observation of the whole period. The distribution is then divided into ten deciles. Therefore, a darker bar in Figure \ref{fig:SectorsClust} means that the average clustering of a sector in a specific year is characterized by a higher value with respect to the clustering coefficients observed in the whole time period considered.
In this way, we take into account also the different financial conditions observed over time. \\
Although in Figure \ref{fig:Sectors} important differences have been noticed between sectors in terms of interconnections with a prevalence of IN and FI sectors, important co-movements across stock markets can be emphasized. It is indeed noticeable that in period of crisis all sectors are characterized by a significant increase of clustering and by a reduction of diversification benefits. \\

\subsection{The effect of alternative estimations of the correlation matrices}

We developed numerical analysis in previous subsections considering correlation matrices estimated via sample approach. However, as shown in \cite{Michaud}, the sample correlation matrix can work poorly in large dimensions for out-of-sample purposes. Although the topic of this paper is not strictly related to the identification of an optimal portfolio, we tested the effect of various estimation methods on the clustering coefficients. \\
In particular, we focus on three alternative approaches to account for the statistical noise. Following \cite{Engle}, we obtained the covariance matrix by applying alternatively the linear shrinkage approach proposed in \cite{LEDOIT2004} and  the non-linear shrinkage approach based on \cite{Ledoit2012}, \cite{Ledoit2015} and \cite{ledoit2016numerical}. 
Additionally, we used the sample estimation and we considered only correlation coefficients that are significantly different from zero at a significance level of 5\%.

\begin{figure}[!htb]
	\centering
	\includegraphics[width=1.0\textwidth]{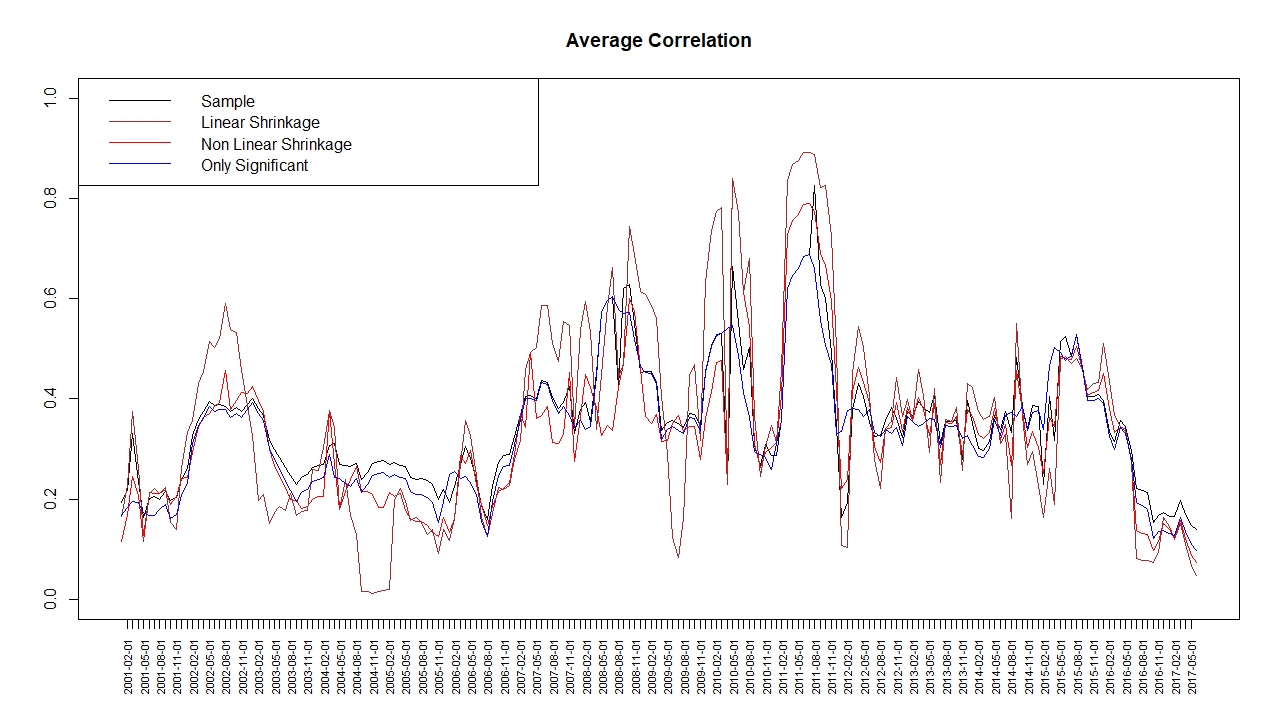}
	\caption{Distributions of elements of the correlation matrix $\boldsymbol{\Rho}_{t}$ for each time period.
	}
	\label{fig:AvCorr}
\end{figure}

\begin{figure}[!htb]
	\centering
	\includegraphics[width=1.0\textwidth]{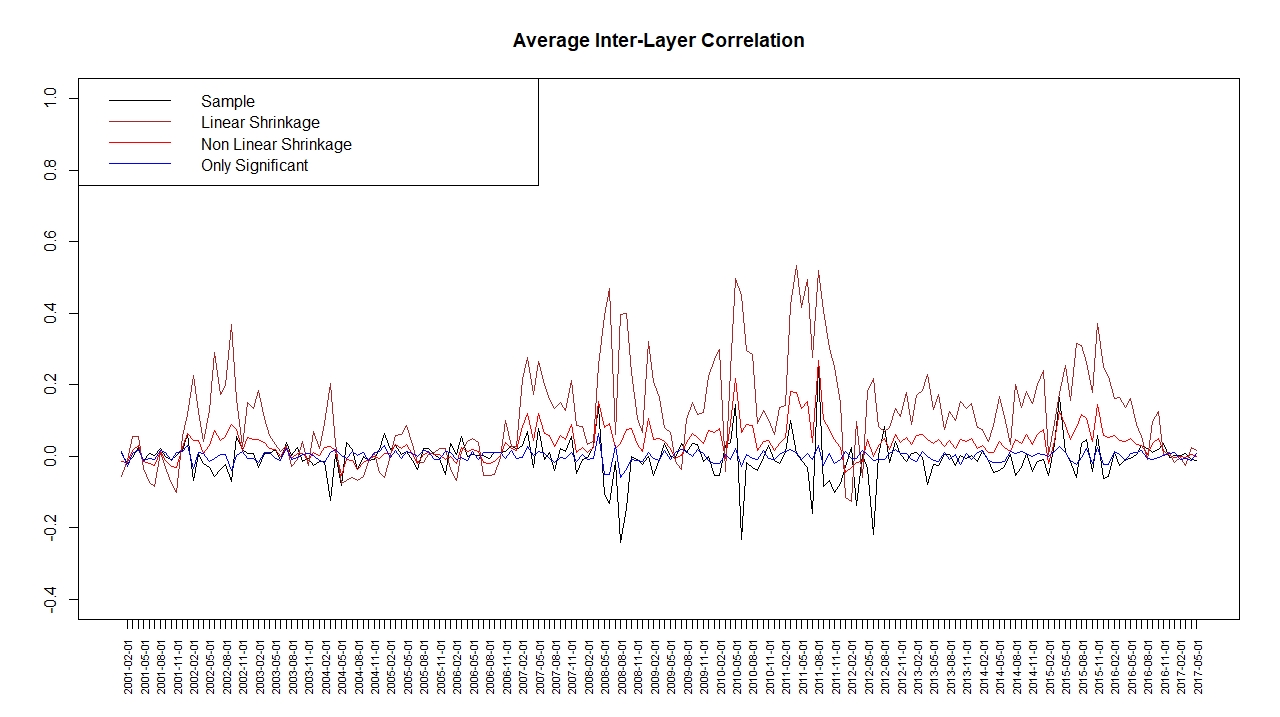}
	\caption{Distributions of elements of the correlation matrix $\boldsymbol{\Rho}_{t,t+1}$ for each couple of subsequent months.
	}
	\label{fig:AvCorrInt}
\end{figure}

To this end, we compare in Figures \ref{fig:AvCorr} and \ref{fig:AvCorrInt} the average correlation coefficients, in each layer and between layers (i.e., the average value of the elements of matrices $\boldsymbol{\Rho}_{t}$ and $\boldsymbol{\Rho}_{t,t+1}$, respectively), obtained by considering alternative approaches for the estimation of the covariance matrix. \\
We notice a similar behaviour over time with a greater volatility of the average correlation when a linear shrinkage estimator is applied.  Also the distributions of coefficients over time are very close to those of Figures \ref{fig:Cor} and \ref{fig:IntCor}.

We repeated the approach proposed in Section \ref{Clustering for multiplex networks} in order to evaluate the effect of the different correlation matrices on the multiplex network and, therefore, on the clustering coefficients.

\begin{figure}[!htb]
	\centering
	\includegraphics[width=12cm, height=6cm]{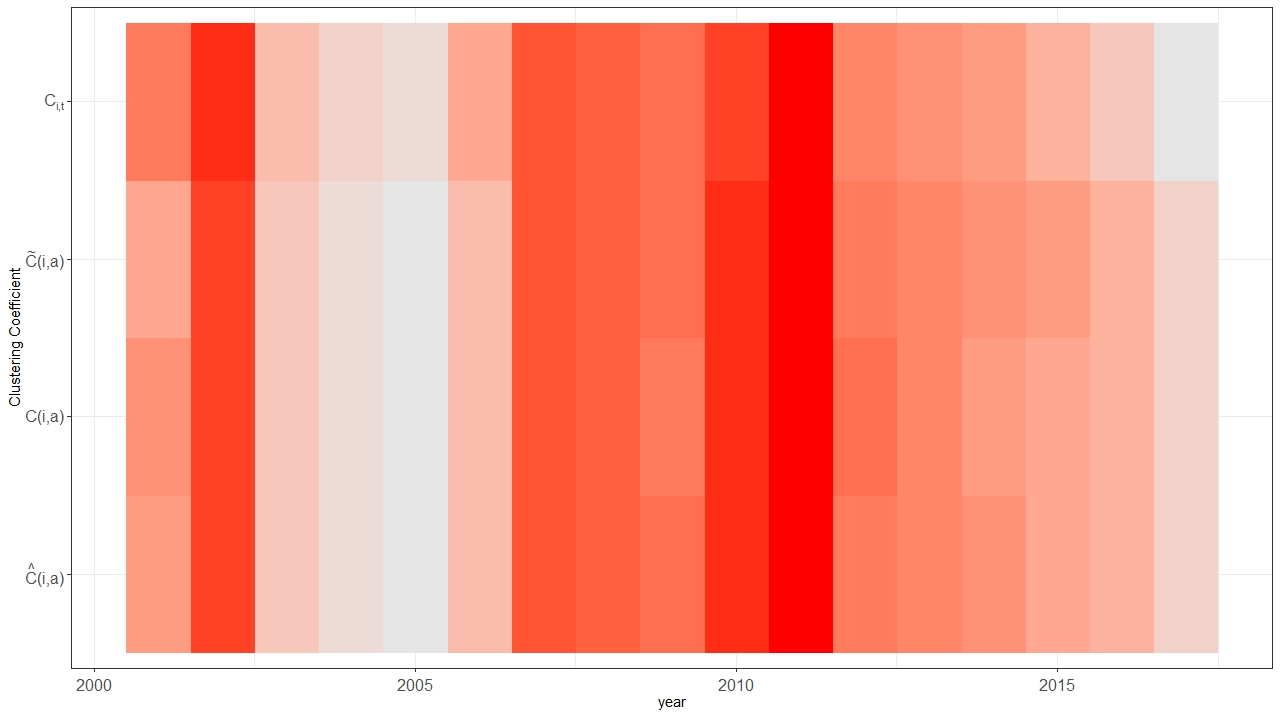}
	\caption{Ranking of average clustering coefficients for each year obtained using correlation matrices estimated via linear shrinkage. Darker bars indicate years characterized by higher clustering coefficients.
	}
	\label{fig:ClustLinear}
\end{figure}

\begin{figure}[!htb]
	\centering
	\includegraphics[width=12cm, height=6cm]{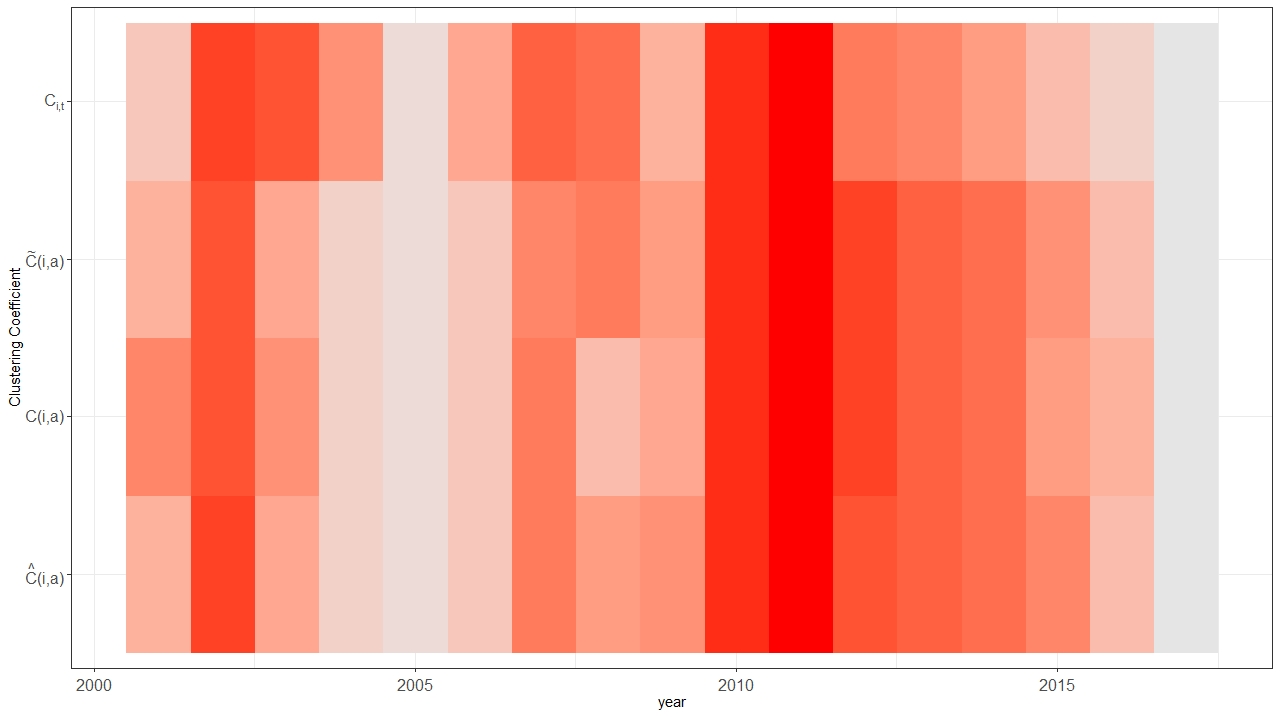}
	\caption{Ranking of average clustering coefficients for each year obtained using correlation matrices estimated via non-linear shrinkage. Darker bars indicate years characterized by higher clustering coefficients.
	}
	\label{fig:ClustNonLinear}
\end{figure}

\begin{figure}[!htb]
	\centering
	\includegraphics[width=12cm, height=6cm]{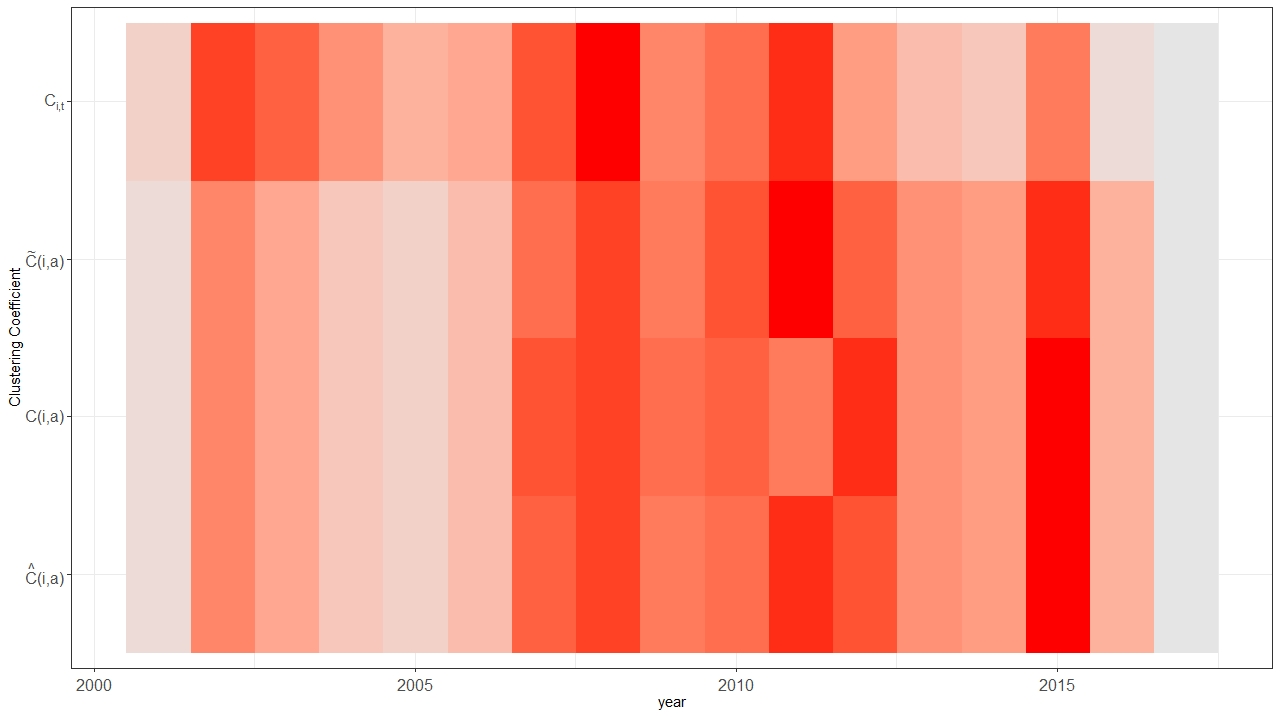}
	\caption{Ranking of average clustering coefficients for each year obtained considering only significant correlation coefficients estimated via sample approach. Darker bars indicate years characterized by higher clustering coefficients.
	}
	\label{fig:SignificantClust}
\end{figure}

In particular, Figures \ref{fig:ClustLinear}, \ref{fig:ClustNonLinear} and \ref{fig:SignificantClust} display the year-based ranking of the alternative clustering coefficients. For the multiplex case, the coefficients have been obtained using networks based on the distance matrices $\mathbf{D}_{t}$ and $\mathbf{D}_{t,t+1}$. These matrices are related to the correlation coefficients estimated through a linear shrinkage, a non-linear shrinkage or considering only significant coefficients of the sample correlation matrix. The high level of similarity in the correlation coefficients derived by the alternative approaches induces that all Figures show a pattern similar to Figure \ref{fig:FigureClust}. In particular, in all cases, it is confirmed a high level of interconnection during periods of financial crises (2007-2008 and 2010-2011). \\ The effect of non-significant correlations is limited. Indeed, Figure \ref{fig:FigureClust}, based on sample estimation, provide the same insights of Figure \ref{fig:SignificantClust} that considers only coefficients significantly different from zero. \\
Focusing on shrinkage methods, strong ties are observed also in 2002 due to the high level of dependence detected by these approaches in that year. \\
Finally, a sector-level analysis has been also developed. Although few differences are noticed, main results observed in Figures \ref{fig:Sectors} and \ref{fig:SectorsClust} are confirmed in all cases.

\newpage
\section{Conclusions}
Complex networks have long gained attention in describing the structure of financial markets. In this paper, we focus on the clustering coefficient, a suitable indicator for catching the level of interconnections in the financial networks. Although several proposals have been provided for the case of monoplex networks, the problem of measuring and assessing local clustering for weighted multiplex network deserves specific attention.
We provide alternative clustering coefficients able to catch the degree of interconnection at both node and layer level, as well as a global version of the clustering coefficient for the whole multiplex network. We highlight how the proposed coefficients are able to catch the effects of both  intra and inter-layer connections.
Additionally, the proposed coefficients represent a generalization of the coefficients already provided in the literature for weighted monolayer 
networks.
The approach has been tested on a large financial temporal multiplex network based on correlation coefficients between returns of assets constituents S\&P 100 index. Results provide additional insights in terms of behaviours of the whole network, specific assets and sectors over time.

\bibliographystyle{model5-names}\biboptions{authoryear}
\bibliographystyle{apa}
\bibliography{Myref}







\end{document}